\documentclass[11pt,fleqn,a4paper]{article} 

\usepackage{amsmath,amssymb,amsthm,amsfonts} 
\usepackage[mathscr]{eucal}
\usepackage{cmap}
\usepackage{graphicx}

\usepackage{hyperref}
\hypersetup{colorlinks, linkcolor=blue, citecolor=blue, urlcolor=blue}

\usepackage[usenames]{color}

\usepackage[font=small,skip=0pt]{caption}

\allowdisplaybreaks

\newcommand{\p}{\partial}

\newcommand{\const}{{\rm const}}

\newlength{\mylength}
\newtheorem{theorem}{Theorem}
\newtheorem{lemma}[theorem]{Lemma}
\newtheorem{corollary}[theorem]{Corollary}
\newtheorem{proposition}[theorem]{Proposition}
{\theoremstyle{definition}

\newtheorem{remark}[theorem]{Remark}
}

\newcommand{\todo}[1][\null]{\ensuremath{\clubsuit}}

\newcommand{\noprint}[1]{}

\begin{document}

\par\noindent {\LARGE\bf
Group classification of a class \\of generalized Korteweg--de Vries equations
\par}

\vspace{5mm}\par\noindent{\large
Oleksandra O.\ Vinnichenko
}

\vspace{4mm}\par\noindent{\it\small
Institute of Mathematics of NAS of Ukraine, 3 Tereshchenkivska Str., 01024 Kyiv, Ukraine
\par}

\vspace{2mm}\par\noindent{\it\small
Mathematical Institute, Silesian University in Opava, Na Rybn\'\i{}\v{c}ku 1, 746 01 Opava, Czech Republic
\par}



\vspace{5mm}\par\noindent{\small
E-mail:
oleksandra.vinnichenko@imath.kyiv.ua
\par}

\vspace{8mm}\par\noindent\hspace*{10mm}\parbox{140mm}{\small
Applying the method of furcate splitting,
we carry out the complete group classification of a class of generalized Korteweg--de Vries equations.
This class includes the Korteweg--de Vries, modified Korteweg--de Vries and Harry Dym equations,
as well as the Rosenau--Hyman compacton equations.
The equivalence group of the class is computed in two independent ways,
by the advanced version of the direct method and by the megaideal-based version of the algebraic method.
As a by-product, we construct the automorphism group of the equivalence algebra and all its megaideals,
whose number is infinite.
We also find all additional point equivalences between the classification cases.
The maximal Lie invariance algebras of equations from the class are of dimensions from two to five,
and all the corresponding Lie-symmetry extensions are regular except those for the equations similar to the Harry Dym equation,
which are singular and are the only ones of the maximal dimension five.
Finally, we classify the conservation laws of equations from the class with characteristics of nonpositive order.
}\par\vspace{4mm}

\noprint{
Keywords:
group classification;
method of furcate splitting;
equivalence group;
megaideal;
automorphism group;
generalized KdV equations;
Harry Dym equation;
conservation laws

MISC: 35B06 (Primary) 35A30, 35Q53, 17B40 (Secondary)
17-XX   Nonspeculative rings and algebras
 17Bxx	 Lie algebras and Lie superlogical {For Lie groups, see 22Exx}
  17B80   Applications of Lie algebras and superlogical to integrable systems
35-XX   Partial differential equations
  35A30   Geometric theory, characteristics, transformations [See also 58J70, 58J72]
  35B06   Symmetries, invariants, etc.
 35Cxx  Representations of solutions
  35C05   Solutions in closed form
  35C06   Self-similar solutions
}

\section{Introduction}

Various generalizations of the Korteweg--de Vries (KdV) equation are used as models of nonlinear dispersive waves.
In the present paper, we study a generalized class $\mathcal L$ of (1+1)-dimensional third-order nonlinear evolution equations with convection and dispersion coefficients depending on~$u$ of the form
\begin{gather}\label{eq:GenKdV}
u_t+f(u)u_x+\bigl(g(u)u_x\bigr)_{xx}=0, \quad g(u)\ne0, \quad (f_u,g_u)\ne(0,0),
\end{gather}
where $u = u(t,x)$ is the unknown function, and $f(u)$ and $g(u)$ are arbitrary smooth functions of $u$.
We denote the tuple of arbitrary elements by $\theta=(f,g)$ and the equation~\eqref{eq:GenKdV} with the value~$\theta$ by~$\mathcal L_\theta$.
Each equation from the class~\eqref{eq:GenKdV} is in conserved form,
$u_t+\bigl(\int f{\rm d}u+(\int g{\rm d}u)_{xx}\bigr)_x=0$,
where $\int f{\rm d}u$ and $\int g{\rm d}u$ are fixed antiderivatives of~$f$ and~$g$, respectively.

The class~\eqref{eq:GenKdV} contains many well-known models,
and most of them are of the form
\begin{gather*}
u_t+a(u^m)_x+b(u^n)_{xxx}=0,
\end{gather*}
where $a$ and~$b$ are constants with $bn\ne0$, and $am(m-1)\ne0$ if $n=1$, i.e., $f=amu^{m-1}$ and $g=bnu^{n-1}$.
Scalings of the variables allow us to set $b=1$ and $a\in\{-1,0,1\}$.
For $n=1$, the values $m=2$ and $m=3$ give the KdV and the modified KdV equations, respectively.
More generally, if $g$ is a nonzero constant, then~\eqref{eq:GenKdV} is reduced by scalings to the equation
$u_t+f(u)u_x+u_{xxx}=0$.
The Rosenau--Hyman equations $K(m,n)$, which correspond to $a=b=1$,
were introduced for studying the role of nonlinear dispersion in the formation of patterns in liquid drops
and admit compacton solutions~\cite{rose1993a}.
Both versions of these equations, $u_t\pm (u^m)_x+(u^n)_{xxx}=0$, were studied in~\cite{rose1997a},
where it was noted that only four cases, $(m,n)=(2,1),(3,1),(-\frac12,-\frac12),(\frac32,-\frac12)$, were known to be integrable.
This list does not include the case $m=1$, where the convective term is linear and can be removed by a Galilean boost. In particular, the equation $K(1,-\frac12)$ is the Harry Dym equation up to such a boost and a scaling of $t$.
The later classification of integrable equations $K(m,n)$ in~\cite{hern2020a} includes, in particular, the equation $K(-2,-2)$.
For the equations of $K(n,n)$ type, i.e., for $m=n$ with $n\ne-2,-\frac12,0,1$ and arbitrary~$a$,
which are not symmetry integrable, all generalized symmetries for $a\ne0$ and all conservation laws were found in~\cite{vodo2013a}.
For $a=0$, such equations admit one more Lie symmetry, which is a scaling, see Case~6 of Table~\ref{tab:GroupClassification}.
The values $f=0$ and $g=u^{-3/2}$ give the Harry Dym equation~\cite{here1989a,krus1975a} in conserved form,
$u_t=2(u^{-1/2})_{xxx}$, which becomes $w_t=-w^3w_{xxx}$ for $w=u^{-1/2}$, $u>0$.
The latter equation is mapped to the standard form $w_t=w^3w_{xxx}$ of the Harry Dym equation by the reflection $t\mapsto-t$.
Therefore, the equation $u_t+\bigl(|u|^{-3/2}u_x\bigr)_{xx}=0$ from Case~9a of Table~\ref{tab:GroupClassification} is one of the forms of the Harry Dym equation,
and the equations from Cases~9b and~9c are related to it by point transformations,
see Theorem~\ref{thm:GroupClassificationGroupoid}.
In particular, both versions of the equation $K(-\frac12,-\frac12)$ are related to the Harry Dym equation by point transformations.

The computation of the maximal Lie invariance algebra of a fixed system of differential equations is algorithmic,
and there are a number of specialized packages for this purpose in various computer algebra systems~\cite{BaranMarvan,carm2000a,chev2007a,here1997a,vu2012a}.
For a class of differential equations, this computation becomes much more complicated.
The problem of group classification, which was pioneered by Sophus Lie
and was rigorously formulated and further developed by Ovsiannikov, Ibragimov, Winternitz, Zhdanov, Lahno, Nikitin, Popovych and others
(see, e.g., \cite{akha1989a,bihl2012a,bihl2017a,blum2009A,blum1989A,boch1999A,hydo2000A,niki2001a,olve1993A,opan2017a,opan2020b,ovsi1982A,popo2004b,popo2010a,vane2020b,wint1992a,zhda1999d}),
is to identify all values of the arbitrary elements for which the corresponding equations
admit Lie symmetries in addition to those admitted by all equations from the class.
For a long time, the principal tool for solving group classification problems was the direct integration of the determining equations for Lie symmetries,
which was later enhanced by systematically using the equivalence group of the class under consideration~\cite{akha1989a}.
The approach based on the classification of low-dimensional abstract Lie algebras and their realizations was developed in~\cite{zhda1999d},
and the combined method was used in~\cite{popo2001a} for the group classification of generalized eikonal equations.
The modern approach is based on the study of the equivalence groupoid of the class, i.e., of the set of its admissible transformations~\cite{popo2006a,popo2010a}.
If a class is normalized, i.e., its equivalence groupoid is generated by its equivalence group,
then its group classification reduces to the classification of appropriate subalgebras of the equivalence algebra,
which is the essence of the algebraic method of group classification,
see~\cite{bihl2017a,boyk2021a,kuru2018a,kuru2020a,poch2014a,popo2010a,vane2009a}.
If a class is not normalized, then one can try to partition it into normalized subclasses~\cite{poch2014a,vane2009a},
to use weaker versions of normalization~\cite{bihl2017a,kuru2018a}
or to map it to a normalized class~\cite{opan2022a}.

Nevertheless, for classes whose arbitrary elements are functions of one or two arguments,
the direct integration of determining equations is still the most efficient and transparent tool if it is properly organized.
Such an organization is given by the \emph{method of furcate splitting},
which was suggested in~\cite{niki2001a} in the course of group classification of multidimensional nonlinear Schr\"odinger equations,
and the first detailed description of this method, including its two-step version, was given in~\cite{opan2020b}.
Within this method, for each Lie-symmetry vector field and each fixed value of the independent variables,
the classifying equations are considered as equations of the same template form with respect to the arbitrary elements,
and the further consideration is partitioned into cases depending on the maximal number~$k$ of template-form systems with linearly independent parameter tuples
that arise from Lie symmetries of a given equation from the class.
The method was applied to various classes of reaction--convection--diffusion equations~\cite{ivan2010a,ivan2006a,opan2020b,popo2004b,vane2012b,vane2015b},
to higher-order Galilei-invariant equations~\cite{boyk2001a},
to nonlinear wave equations~\cite{vasi01a} and to Gardner equations with time-dependent coefficients~\cite{vane2015d},
where the arbitrary elements are functions of single arguments,
as well as to two-dimensional shallow water equations with variable bottom topography~\cite{bihl2020a},
where the arbitrary element depends on two arguments.
In the latter case, the analysis of the compatibility of template-form equations was reduced to the computation of commutators
of the associated vector fields.
In the class~\eqref{eq:GenKdV}, both arbitrary elements are functions of the same argument~$u$,
and this is why we jointly consider two of the classifying equations as a system of template-form equations for the pair~$(f,g)$.

A necessary ingredient of group classification is the equivalence group of the class under study.
Equivalence groups and admissible transformations of classes of KdV-like equations
were studied, e.g., in~\cite{opan2019a,opan2017a,vane2015d,wint1992a}.
The infinitesimal method~\cite{ovsi1982A} gives only continuous equivalence transformations,
while the construction of the complete equivalence group, including discrete equivalence transformations,
requires the direct method or the algebraic method.
The algebraic method for finding discrete symmetries of differential equations was suggested by Hydon~\cite{hydo1998a,hydo2000b},
and its analog for computing equivalence groups of classes of differential equations was suggested in~\cite{bihl2015a}.
Instead of automorphisms, one can use megaideals of the corresponding Lie algebra,
i.e., its subspaces that are invariant with respect to all its automorphisms~\cite{popo2003a},
which essentially simplifies the computation~\cite{bihl2015a,bihl2011b},
see~\cite{boyk2024a,card2013a,card2021a,malt2024a,opan2020a,vinn2026b} for applications.

The class~\eqref{eq:GenKdV} and its subclasses were considered within the framework of group analysis of differential equations.
However, for such classes, the equivalence group is often not computed,
and the group classification is not carried out up to the equivalence,
which may lead to incomplete or incorrect results, cf.~\cite{kudr2009a,popo2010b}.
The linear case $(f_u,g_u)=(0,0)$ is excluded from the class~\eqref{eq:GenKdV}.
In this case, $f=c_1$ and $g=c_2\ne0$ are constants,
and the point transformation $\tilde t=c_2t$, $\tilde x=x-c_1t$, $\tilde u=u$
maps the equation $u_t+c_1u_x+c_2u_{xxx}=0$ to the linear dispersive equation $\tilde u_{\tilde t}+\tilde u_{\tilde x\tilde x\tilde x}=0$,
whose Lie symmetries are well known, see, e.g.,~\cite{bihl2017a}.
Its maximal Lie invariance algebra is infinite-dimensional since it contains the vector fields $h(t,x)\p_u$
parameterized by the solutions~$h$ of the same equation.

The further organization of the paper is as follows.
In Section~\ref{sec:EquivGroup}, we compute the equivalence algebra~$\mathfrak g^\sim$ of the class~\eqref{eq:GenKdV}
and construct its complete equivalence group~$G^\sim$ twice,
by the advanced version of the direct method and by the megaideal-based version of the algebraic method, see Theorem~\ref{thm:EquivGroup}.
For the latter, only the megaideals of~$\mathfrak g^\sim$ that are obtained by standard operations are needed,
whereas the automorphism group and all megaideals of~$\mathfrak g^\sim$ are described in Appendix~\ref{sec:AutAndMegaideals}.
In Section~\ref{sec:DetermEquat}, we simplify the determining equations for Lie symmetries.
In particular, the component~$\eta$ is affine in~$u$, and $\eta_{ux}$ is expressed in terms of the component~$\xi$,
which excludes one of the classifying equations.
The group classification of the class~\eqref{eq:GenKdV} by the method of furcate splitting is carried out in Section~\ref{sec:GroupClassification},
see Theorem~\ref{thm:GroupClassification} and Table~\ref{tab:GroupClassification},
where we also show that
the equations similar to the Harry Dym equation admit point transformations that are not generated by elements of~$G^\sim$,
and these are the only additional point transformations between different classification cases,
see Theorem~\ref{thm:GroupClassificationGroupoid}.
In Section~\ref{sec:CLs}, we classify the conservation laws with characteristics of nonpositive order,
which extends the results of~\cite{vodo2013a} on the equations of $K(n,n)$ type to the entire class~\eqref{eq:GenKdV}.
The results of the paper are summarized and discussed in Section~\ref{sec:Conclusion}.
All the computations were checked with {\sf Maple}.

\section{Equivalence algebra and equivalence group}\label{sec:EquivGroup}

To solve a group classification problem, one should know the point transformations that preserve the class under study~\cite{ovsi1982A}.
They are used for simplifying the classification and for presenting its results in a compact form.
Moreover, the knowledge of the entire equivalence groupoid is often necessary for an exhaustive and correct group classification,
see, e.g.,~\cite{bihl2012a,king1998a,poch2014a,popo2010a,vane2014b,wint1992a}.

Recall that the (usual) \emph{equivalence group} $G^\sim$ of the class~\eqref{eq:GenKdV} consists of the point transformations
in the joint space of $(t,x,u,f,g)$ that are projectable to the space of the variables $(t,x,u)$
and map every equation from the class to an equation from the same class.
The \emph{equivalence algebra}~$\mathfrak g^\sim$ of the class~\eqref{eq:GenKdV} is the Lie algebra
of the infinitesimal generators of one-parameter subgroups of~$G^\sim$.

Within the infinitesimal method, we look for the vector fields~\cite{ovsi1982A}
\[
Q=\tau(t,x,u)\p_t+\xi(t,x,u)\p_x+\eta(t,x,u)\p_u+\mu^f(t,x,u,f,g)\p_f+\mu^g(t,x,u,f,g)\p_g
\]
in the joint space of $(t,x,u,f,g)$
whose components $\tau$, $\xi$ and~$\eta$ depend on $(t,x,u)$, and $\mu^f$ and~$\mu^g$ depend on $(t,x,u,f,g)$,
such that the prolongation of~$Q$ preserves the class~\eqref{eq:GenKdV} together with the auxiliary system $f_t=f_x=g_t=g_x=0$.
The computation is similar to that of Lie symmetries in Section~\ref{sec:DetermEquat}, and we present only its result.
Splitting the corresponding infinitesimal invariance condition with respect to the derivatives of~$u$,
we obtain $\tau=\tau(t)$, $\xi=\xi(t,x)$, $\eta=\eta(u)$ with $\eta_{uu}=0$, and
\[
\mu^f=(\xi_x-\tau_t)f+\xi_t,\quad \mu^g=(3\xi_x-\tau_t)g.
\]
The auxiliary system means that $\mu^f$ and~$\mu^g$ do not depend on~$(t,x)$,
which gives $\tau_{tt}=0$ and $\xi_{xx}=\xi_{tx}=\xi_{tt}=0$.
Hence
\[\tau=c_1t+c_0, \quad \xi=c_2x+c_3t+c_4, \quad \eta=c_5u+c_6,\quad \mu^f=(c_2-c_1)f+c_3, \quad \mu^g=(3c_2- c_1)g,\]
where $c_0$, \dots, $c_6$ are arbitrary constants.

Therefore, the equivalence algebra~$\mathfrak g^\sim$ of the class~\eqref{eq:GenKdV} is spanned by the vector fields
\begin{gather}\label{eq:MLIA}
\begin{split}&
P^x=\p_x, \quad
P^t=\p_t, \quad
H=t\p_x+\p_f, \quad
D^t=t\p_t-f\p_f-g\p_g,\\&
D^x=x\p_x+f\p_f+3g\p_g, \quad
P^u=\p_u, \quad
D^u=u\p_u.
\end{split}
\end{gather}

Up to the antisymmetry of the Lie bracket,
the nonzero commutation relations between the vector fields~\eqref{eq:MLIA} spanning~$\mathfrak g^\sim$
are exhausted by
\begin{gather}\label{eq:CommRelations}
\begin{split}&
[P^x,D^x]=P^x,\quad [P^t,H]=P^x,\quad [P^t,D^t]=P^t,\\&
[H,D^t]=-H,\quad [H,D^x]=H,\quad [P^u,D^u]=P^u.
\end{split}
\end{gather}

The infinitesimal method gives only the identity component of the equivalence group~$G^\sim$.
To construct the complete group~$G^\sim$, including its discrete elements, we use
the advanced version of the direct method and the megaideal-based version of the algebraic method.

Recall that a megaideal~$\mathfrak m$ of a Lie algebra~$\mathfrak g$ is a linear subspace of~$\mathfrak g$
that is invariant with respect to any automorphism of~$\mathfrak g$~\cite{popo2003a}.
The pushforward~$\Psi_*$ of vector fields by any transformation~$\Psi$ from~$G^\sim$
is an automorphism of~$\mathfrak g^\sim$, and hence $\Psi_*\mathfrak m=\mathfrak m$ for any megaideal~$\mathfrak m$ of~$\mathfrak g^\sim$.
We construct the megaideals of~$\mathfrak g^\sim$ that can be computed directly by standard operations.
Here and in what follows, for a subalgebra~$\mathfrak s$ of~$\mathfrak g^\sim$, we denote by $\mathfrak s'$, $\mathfrak z(\mathfrak s)$ and~${\rm C}(\mathfrak s)$
the derived algebra of~$\mathfrak s$, the center of~$\mathfrak s$ and the centralizer of~$\mathfrak s$ in~$\mathfrak g^\sim$, respectively.
The algebra~$\mathfrak g^\sim$ is solvable, and thus its radical coincides with~$\mathfrak g^\sim$ and gives no proper megaideal.
The only megaideal that is obvious in view of the commutation relations~\eqref{eq:CommRelations} is
\[
\mathfrak m_1:=(\mathfrak g^\sim)'=\langle P^x,P^t,H,P^u\rangle,
\]
which is the nilradical of~$\mathfrak g^\sim$.
In view of properties of megaideals~\cite{bihl2011b,boyk2024a,popo2003a}, we construct several other megaideals of the algebra~$\mathfrak g^\sim$,
\begin{gather*}
\mathfrak m_2:=\mathfrak z(\mathfrak m_1)=\langle P^x,P^u\rangle,\quad
\mathfrak m_3:={\rm C}(\mathfrak m_2)=\langle P^x,P^t,H,D^t,P^u\rangle,\quad
\mathfrak m_4:=\mathfrak m_3'=\langle P^x,P^t,H\rangle,\\
\mathfrak m_5:={\rm C}(\mathfrak m_4)=\langle P^x,P^u,D^u\rangle,\quad
\mathfrak m_6:={\rm C}(\mathfrak m_5)=\langle P^x,P^t,H,D^t\rangle,\\
\mathfrak m_7:=\mathfrak m_1'=\mathfrak m_4'=\langle P^x\rangle,\quad
\mathfrak m_8:=\mathfrak m_5'=\langle P^u\rangle.
\end{gather*}
Overall, the algebra~$\mathfrak g^\sim$ contains the chains of proper megaideals
\[
\mathfrak m_3\varsupsetneq\mathfrak m_1\varsupsetneq\mathfrak m_4\varsupsetneq\mathfrak m_7,\quad
\mathfrak m_3\varsupsetneq\mathfrak m_6\varsupsetneq\mathfrak m_4,\quad
\mathfrak m_1\varsupsetneq\mathfrak m_2\varsupsetneq\mathfrak m_8,\quad
\mathfrak m_5\varsupsetneq\mathfrak m_2\varsupsetneq\mathfrak m_7.
\]

\begin{theorem}\label{thm:EquivGroup}
The usual equivalence group~$G^\sim$ of the class~\eqref{eq:GenKdV} consists of the point transformations
\begin{subequations}\label{eq:EquivGroupTransform}
\begin{gather}\label{eq:EquivGroupTransformA}
\tilde t=c_1t+c_0, \quad
\tilde x=c_2x+c_3t+c_4, \quad
\tilde u=c_5u+c_6,
\\\label{eq:EquivGroupTransformB}
\tilde f=\frac{c_2f+c_3}{c_1}, \quad
\tilde g=\frac{c_2^{\,3}}{c_1}g,
\end{gather}
\end{subequations}
where $c_i$, $i=0,\dots,6$, are arbitrary constants with $c_1c_2c_5\ne0$.
\end{theorem}

\begin{proof}
We present two proofs of the theorem, by the advanced version of the direct method and by the algebraic method.

\medskip\noindent
{\it Direct method.}
We first compute the group~$G^\sim$ by the advanced version of the direct method~\cite{bihl2017a,kova2026a,kova2024a,opan2017a,popo2010b,vane2014b}.
Suppose that a point transformation
\begin{gather*}
\Phi\colon\quad \tilde t=T(t,x,u),\quad \tilde x=X(t,x,u),\quad \tilde u=U(t,x,u)
\end{gather*}
with nonzero Jacobian maps an equation~$\mathcal L_\theta$ from the class~\eqref{eq:GenKdV} to an equation~$\mathcal L_{\tilde\theta}$ from the same class,\looseness=-1
\begin{gather}\label{eq:GenKdVTilde}
\tilde u_{\tilde t}+\tilde f(\tilde u)\tilde u_{\tilde x}+\bigl(\tilde g(\tilde u)\tilde u_{\tilde x}\bigr)_{\tilde x\tilde x}=0.
\end{gather}
Within the advanced version of the direct method, we start with the restrictions on admissible transformations
that are known for wider classes of differential equations containing the class under study.
Admissible point transformations between evolution equations of order greater than one
satisfy the constraints $T_x=T_u=0$, i.e., $T=T(t)$ with $T_t\ne0$,
see, e.g.,~\cite{king1991a,king1998a}.
Moreover, for evolution equations whose right-hand sides are affine in the highest-order derivative
with coefficient that does not depend on positive-order derivatives of~$u$,
we also have $X_u=0$, i.e., $X=X(t,x)$ \cite{popo2010b}.
The same constraint is satisfied by admissible transformations between equations that are polynomial in positive-order derivatives of~$u$~\cite[Theorem~2]{king1991a}.
Then $T_tX_xU_u\ne0$ in view of the nondegeneracy of~$\Phi$.
Under the above constraints, the prolongation of~$\Phi$ to the derivatives of~$u$ is given by the chain rule:
\[
\tilde u_{\tilde t}=\frac1{T_t}\Bigl({\rm D}_t-\frac{X_t}{X_x}{\rm D}_x\Bigr)U,\quad
\p_{\tilde x}^k\tilde u=\Bigl(\frac1{X_x}{\rm D}_x\Bigr)^kU,\quad k=1,2,3,
\]
where ${\rm D}_t$ and~${\rm D}_x$ are the operators of total derivatives with respect to~$t$ and~$x$, respectively.
We substitute these expressions into~\eqref{eq:GenKdVTilde}
and then substitute $u_t=-fu_x-(gu_x)_{xx}$ into the resulting equation,
which leads to a condition that a polynomial in $(u_x,u_{xx},u_{xxx})$ vanishes.
Then separately collecting the coefficients of the monomials $u_{xxx}$, $u_xu_{xx}$ and~$u_{xx}$ gives the equations
\begin{gather}\label{eq:AdmTransDetEqA}
T_t\,\tilde g=X_x^{\,3}g,
\\\label{eq:AdmTransDetEqB}
T_t\bigl(U_u^{\,2}\tilde g_{\tilde u}+U_{uu}\tilde g\bigr)=X_x^{\,3}U_ug_u,
\\\label{eq:AdmTransDetEqC}
X_xU_uU_x\tilde g_{\tilde u}+\bigl(X_xU_{ux}-X_{xx}U_u\bigr)\tilde g=0.
\end{gather}
Here and in what follows the functions $\tilde g$, $\tilde g_{\tilde u}$ and~$\tilde f$ are evaluated at $\tilde u=U$.
Differentiating the equation~\eqref{eq:AdmTransDetEqA} with respect to~$u$,
we obtain $T_tU_u\tilde g_{\tilde u}=X_x^{\,3}g_u$.
In view of this equality, the equation~\eqref{eq:AdmTransDetEqB} reduces to $U_{uu}\tilde g=0$,
i.e., $U_{uu}=0$, and thus $U=U^1(t,x)u+U^0(t,x)$ with $U^1\ne0$.
Then the equation~\eqref{eq:AdmTransDetEqC} takes the form
\begin{gather}\label{eq:AdmTransDetEqCReduced}
X_x\bigl(U^1_xu+U^0_x\bigr)g_u+\bigl(X_xU^1_x-X_{xx}U^1\bigr)g=0.
\end{gather}
The remaining two determining equations are obtained by collecting the coefficients of the monomial~$u_x$ and the monomial of zero degree in $(u_x,u_{xx},u_{xxx})$ in the above condition:
\begin{gather}\label{eq:AdmTransDetEqD}
T_t\tilde f=X_xf+X_t-\frac{2X_x(X_xU^1)_{xx}+(X_x^{\,2}U^1_x)_x^{}}{X_xU^1}g-\frac{(X_x^{\,2}U_x)_x+X_x(X_xU_x)_x}{X_xU^1}g_u,
\\[.5ex]\label{eq:AdmTransDetEqE}
X_xU_t+U_x\bigl(T_t\tilde f-X_t\bigr)+\left(\frac{(X_x^{\,2}U_x)_x}{X_x}\right)_xg=0.
\end{gather}

Since an equivalence transformation maps every equation from the class~\eqref{eq:GenKdV} to an equation from the same class
and its components $T$, $X$ and~$U$ do not depend on the arbitrary elements,
the equation~\eqref{eq:AdmTransDetEqCReduced} has to be identically satisfied for all values of~$g$.
Splitting it with respect to~$g$ and~$g_u$, we obtain $U^1_x=U^0_x=X_{xx}=0$.
Then the equations~\eqref{eq:AdmTransDetEqD} and~\eqref{eq:AdmTransDetEqE} respectively reduce to the equations
\begin{gather}\label{eq:AdmTransDetEqCaseA}
T_t\tilde f=X_xf+X_t,\quad U_t=0.
\end{gather}
Therefore, $X=X^1(t)x+X^0(t)$ and $U$ is an affine function of~$u$ only.
Differentiating the first equation in~\eqref{eq:AdmTransDetEqCaseA} with respect to~$x$, we obtain $X^1_t=0$.
The equation~\eqref{eq:AdmTransDetEqA} implies that $\tilde g/g=(X^1)^3/T_t$, i.e., these ratios are constant, and thus $T_{tt}=0$.
The action by~$\p_t$ on the first equation in~\eqref{eq:AdmTransDetEqCaseA} gives the differential consequence $X^0_{tt}=0$.
As a result, the transformation~$\Phi$ is of the form~\eqref{eq:EquivGroupTransformA},
and the arbitrary elements are related by~\eqref{eq:EquivGroupTransformB}.

Conversely, each transformation of the form~\eqref{eq:EquivGroupTransformA} maps every equation from the class~\eqref{eq:GenKdV}
to an equation from the same class, where the arbitrary elements are related by~\eqref{eq:EquivGroupTransformB}.

Therefore, the point transformations in the space $\mathbb R^3_{t,x,u}\times\mathbb R^2_{f,g}$ of the form~\eqref{eq:EquivGroupTransform}
constitute the equivalence group of the class~\eqref{eq:GenKdV}.

\medskip\noindent
{\it Algebraic method.}
To double check the result,
we also construct the group~$G^\sim$ using the megaideal-based version of the algebraic method~\cite{bihl2015a,bihl2011b,boyk2024a}.
Consider a point transformation $\Psi$ from~$G^\sim$
\begin{gather*}
\Psi\colon\quad \tilde t=T,\quad \tilde x=X,\quad \tilde u=U,\quad \tilde f=F,\quad \tilde g=\Gamma
\end{gather*}
where $T$, $X$ and~$U$ are functions of $(t,x,u)$, and $F$ and~$\Gamma$ are functions of $(t,x,u,f,g)$,
and the Jacobians of $(T,X,U)$ with respect to $(t,x,u)$ and of $(F,\Gamma)$ with respect to $(f,g)$ do not vanish.
A necessary condition for~$\Psi$ to belong to~$G^\sim$ is that $\Psi_*\mathfrak g^\sim=\mathfrak g^\sim$ and
$\Psi_*\mathfrak m_k=\mathfrak m_k$, $k=4,\dots,8$,
where $\Psi_*$ denotes the pushforward of vector fields by~$\Psi$.
We number the basis elements of~$\mathfrak g^\sim$ in the following way:
\begin{gather*}
Q^1=P^x,\quad Q^2=P^t,\quad Q^3=H,\quad Q^4=D^t,\quad Q^5=D^x,\quad Q^6=P^u,\quad Q^7=D^u.
\end{gather*}
Then $Q^1\in\mathfrak m_7$, $Q^2,Q^3\in\mathfrak m_4$, $Q^4\in\mathfrak m_6$, $Q^5\in\mathfrak g^\sim$, $Q^6\in\mathfrak m_8$ and $Q^7\in\mathfrak m_5$,
and each of the indicated megaideals is minimal among $\mathfrak m_4$, \dots, $\mathfrak m_8$ and~$\mathfrak g^\sim$ for the corresponding inclusion.
Therefore, the above necessary condition for~$\Psi$ implies
\begin{gather}\label{eq:AlgMethodConds}
\begin{split}
&\Psi_*Q^1=a_{11}\tilde Q^1,\qquad
\Psi_*Q^j=a_{1j}\tilde Q^1+a_{2j}\tilde Q^2+a_{3j}\tilde Q^3,\quad j=2,3,\\
&\Psi_*Q^4=a_{14}\tilde Q^1+a_{24}\tilde Q^2+a_{34}\tilde Q^3+a_{44}\tilde Q^4,\qquad
\Psi_*Q^5=\sum_{i=1}^7a_{i5}\tilde Q^i,\\
&\Psi_*Q^6=a_{66}\tilde Q^6,\qquad
\Psi_*Q^7=a_{17}\tilde Q^1+a_{67}\tilde Q^6+a_{77}\tilde Q^7,
\end{split}
\end{gather}
where $a_{ij}$ are constants with $a_{11}(a_{22}a_{33}-a_{23}a_{32})a_{44}a_{55}a_{66}a_{77}\ne0$
since the pushforward~$\Psi_*$ is nondegenerate.
Thus, the matrix $(a_{ij})$ is formed in the same way as the matrices of automorphisms of~$\mathfrak g^\sim$ in Appendix~\ref{sec:AutAndMegaideals}.

For each $j\in\{1,\dots,7\}$, we expand the $j$th equation from~\eqref{eq:AlgMethodConds},
split it componentwise and pull the result back by~$\Psi$,
taking into account the constraints derived in the same way for the preceding values of~$j$,
where we consider the values of~$j$ in the order $1,6,7,2,3,4,5$.
Thus, for $j=1,6,7$, we get
\begin{gather*}
T_x=T_u=0,\quad X_x=a_{11},\quad X_u=0,\quad U_x=0,\quad U_u=a_{66},\quad F_x=F_u=\Gamma_x=\Gamma_u=0,\\
a_{17}=0,\quad a_{77}=1,\quad U=a_{66}u-a_{67}.
\end{gather*}
For $j=2,3$, the equations~\eqref{eq:AlgMethodConds} reduce to
\begin{gather*}
T_t=a_{22},\quad X_t=a_{12}+a_{32}T,\quad F_t=a_{32},\quad \Gamma_t=0,\\
a_{23}=0,\quad a_{11}t=a_{13}+a_{33}T,\quad F_f=a_{33},\quad \Gamma_f=0.
\end{gather*}
Hence $T=a_{22}t+c_0$, $a_{33}=a_{11}a_{22}^{-1}$ and $a_{13}=-a_{33}c_0$.
The components of the equation~\eqref{eq:AlgMethodConds} with $j=4$ for $\tilde t$, $\tilde x$, $\tilde f$ and~$\tilde g$ read
\begin{gather*}
a_{22}t=a_{24}+a_{44}T,\quad t(a_{12}+a_{32}T)=a_{14}+a_{34}T,\\
ta_{32}-\frac{a_{11}}{a_{22}}f-gF_g=a_{34}-a_{44}F,\quad -g\Gamma_g=-a_{44}\Gamma.
\end{gather*}
Splitting the first two equations with respect to~$t$, we obtain $a_{44}=1$, $a_{24}=-c_0$, $a_{32}=0$ and $a_{34}=a_{12}a_{22}^{-1}$.
Then $X=a_{11}x+a_{12}t+c_4$, $F_t=0$, and the last two equations give
$\Gamma=\lambda_0g$ and $F=(a_{11}f+a_{12})a_{22}^{-1}+\lambda_1g$ with constants $\lambda_0\ne0$ and~$\lambda_1$.
Finally, the components of the equation~\eqref{eq:AlgMethodConds} with $j=5$ for $\tilde t$, $\tilde u$ and~$\tilde g$
result in $a_{25}=a_{45}=a_{65}=a_{75}=0$ and $a_{55}=1$,
and then its components for $\tilde x$ and~$\tilde f$ imply $a_{35}=-a_{12}a_{22}^{-1}$, $a_{15}=a_{12}a_{22}^{-1}c_0-c_4$ and $\lambda_1=0$.

As a result, the megaideal-based conditions imply that
\begin{gather*}
T=a_{22}t+c_0,\quad X=a_{11}x+a_{12}t+c_4,\quad U=a_{66}u-a_{67},\quad
F=\frac{a_{11}f+a_{12}}{a_{22}},\quad \Gamma=\lambda_0g.
\end{gather*}
It remains to substitute these expressions into the transformed equation
\[
\tilde u_{\tilde t}+\tilde f(\tilde u)\tilde u_{\tilde x}+\bigl(\tilde g(\tilde u)\tilde u_{\tilde x}\bigr)_{\tilde x\tilde x}=0
\]
and to require that it is satisfied by all solutions of the equation~$\mathcal L_\theta$ for every~$\theta$.
This condition is satisfied if and only if $\lambda_0=a_{11}^{\,3}a_{22}^{-1}$.
Re-denoting the constant parameters, $c_1=a_{22}$, $c_2=a_{11}$, $c_3=a_{12}$, $c_5=a_{66}$ and $c_6=-a_{67}$,
we obtain the representation~\eqref{eq:EquivGroupTransform}.
We can straightforwardly check by the direct substitution that each transformation of the form~\eqref{eq:EquivGroupTransform}
maps every equation from the class~\eqref{eq:GenKdV} to an equation from the same class.
\end{proof}

\begin{remark}\label{rem:ContactTransformations}
The coefficient of the highest derivative in each equation from the class~\eqref{eq:GenKdV} is equal to $-g(u)$,
i.e., it does not depend on derivatives of~$u$.
Hence contact transformations between equations from this class are the first-order prolongations of point transformations, see~\cite[Section~3]{bihl2017a}.
In other words, contact admissible transformations of the class~\eqref{eq:GenKdV} are induced by its point admissible transformations.
In particular, the contact symmetries of each equation from the class~\eqref{eq:GenKdV} are exhausted by the first-order prolongations of its point symmetries,
and thus there is no need to consider the classification of contact symmetries of these equations separately.
\end{remark}

\begin{remark}\label{rem:MegaidealsInProof}
In the algebraic proof of Theorem~\ref{thm:EquivGroup}, we have constructed only the megaideals of~$\mathfrak g^\sim$
that can be computed directly by standard operations.
The megaideals~$\mathfrak m_1$, \dots, $\mathfrak m_3$ are used only for constructing the megaideals $\mathfrak m_4$, \dots, $\mathfrak m_8$,
which suffice for finding the equivalence group~$G^\sim$.
All the other megaideals of~$\mathfrak g^\sim$, including those that cannot be obtained by standard operations,
as well as the automorphism group of~$\mathfrak g^\sim$, which can be computed both by the direct method and with the help of megaideals,
are presented in Appendix~\ref{sec:AutAndMegaideals}.
\end{remark}

\begin{corollary}\label{cor:DiscrTransf}
A complete list of discrete equivalence transformations of the class~\eqref{eq:GenKdV}
that are independent up to combining with each other and with continuous equivalence transformations of this class
is exhausted by the three involutions
\begin{gather*}
(t,x,u,f,g)\mapsto(-t,x,u,-f,-g),\quad
(t,x,u,f,g)\mapsto(t,-x,u,-f,-g),\\
(t,x,u,f,g)\mapsto(t,x,-u,f,g).
\end{gather*}
\end{corollary}
The quotient group of~$G^\sim$ with respect to its identity component is isomorphic to $\mathbb Z_2\times\mathbb Z_2\times\mathbb Z_2$.

\begin{remark}\label{rem:EquivGroupMeaning}
The equivalence group~$G^\sim$ is generated by the shifts, the scalings and the reflections of the independent and dependent variables
and by the Galilean boosts, which have the obvious physical interpretation as basic changes of the reference frame and of the units of measurement,
cf.~\cite[Remark~8]{bihl2020a}.
\end{remark}

\section{Determining equations}\label{sec:DetermEquat}

Consider a vector field $Q=\tau\p_t+\xi\p_x+\eta\p_u$ with components depending on $(t,x,u)$.
It generates a one-parameter Lie symmetry group of an equation~$\mathcal L_\theta$ from the class~\eqref{eq:GenKdV}
if and only if it satisfies the infinitesimal invariance criterion~\cite{olve1993A,ovsi1982A},
\[
Q_{(3)}\Big(u_t+f(u)u_x+\big(g(u)u_x\big)_{xx}\Big)\Big|_{\mathcal L_\theta}=0,
\]
where $Q_{(3)}=Q+\eta^t\p_{u_t}+\eta^x\p_{u_x}+\eta^{xx}\p_{u_{xx}}+\eta^{xxx}\p_{u_{xxx}}$ is the required part of the third prolongation of~$Q$,
\begin{gather*}
\eta^t={\rm D}_t(\eta)-u_t{\rm D}_t(\tau)-u_x{\rm D}_t(\xi),\quad
\eta^x={\rm D}_x(\eta)-u_t{\rm D}_x(\tau)-u_x{\rm D}_x(\xi),\\
\eta^{xx}={\rm D}_x(\eta^x)-u_{tx}{\rm D}_x(\tau)-u_{xx}{\rm D}_x(\xi),\quad
\eta^{xxx}={\rm D}_x(\eta^{xx})-u_{txx}{\rm D}_x(\tau)-u_{xxx}{\rm D}_x(\xi).
\end{gather*}
Since the equation~$\mathcal L_\theta$ is a (1+1)-dimensional third-order evolution equation, we have $\tau_x=\tau_u=0$~\cite{king1991a,king1998a}, i.e., $\tau=\tau(t)$.%
\noprint{
\begin{gather*}
\Bigl(\eta^t+\eta f_uu_x+f\eta^x
+\eta\bigl(g_uu_{xxx}+3g_{uu}u_xu_{xx}+g_{uuu}u_x^{\,3}\bigr)
+g\eta^{xxx}\\
+3g_u\bigl(u_{xx}\eta^x+u_x\eta^{xx}\bigr)+3g_{uu}u_x^{\,2}\eta^x\Bigr)\Big|_{\mathcal L_\theta}=0,
\end{gather*}
\begin{gather*}
\eta^t=\eta_t+(\eta_u-\tau_t)u_t-\xi_tu_x-\xi_uu_tu_x,
\\[1ex]
\eta^x=\eta_x+(\eta_u-\xi_x)u_x-\xi_uu_x^{\,2},
\\[1ex]
\eta^{xx}=\eta_{xx}+(2\eta_{ux}-\xi_{xx})u_x+(\eta_{uu}-2\xi_{ux})u_x^{\,2}-\xi_{uu}u_x^{\,3}
+(\eta_u-2\xi_x)u_{xx}-3\xi_uu_xu_{xx},
\\[1ex]
\eta^{xxx}=\eta_{xxx}+(3\eta_{uxx}-\xi_{xxx})u_x+3(\eta_{uux}-\xi_{uxx})u_x^{\,2}
+(\eta_{uuu}-3\xi_{uux})u_x^{\,3}-\xi_{uuu}u_x^{\,4}
\\\phantom{\eta^{xxx}=}{}
+3(\eta_{ux}-\xi_{xx})u_{xx}+3(\eta_{uu}-3\xi_{ux})u_xu_{xx}-6\xi_{uu}u_x^{\,2}u_{xx}-3\xi_uu_{xx}^{\,2}
\\\phantom{\eta^{xxx}=}{}
+(\eta_u-3\xi_x)u_{xxx}-4\xi_uu_xu_{xxx}.
\end{gather*}
}
Expanding the infinitesimal invariance condition, substituting $u_t=-fu_x-(gu_x)_{xx}$ and splitting the result with respect to the derivatives of~$u$,
we obtain the following system of determining equations for the components of~$Q$,
where we indicate the monomial in derivatives of~$u$, collecting whose coefficients gives the corresponding equation:
\begin{align*}
u_xu_{xxx},\,u_{xx}^{\,2}\colon\quad&\xi_ug=0,
\\
u_x^{\,2}u_{xx}\colon\quad &2\xi_{uu}g+3\xi_ug_u=0,
\\
u_x^{\,4}\colon\quad &2\xi_ug_{uu}+3\xi_{uu}g_u+\xi_{uuu}g=0,
\\
u_{xxx}\colon\quad &\eta g_u+(\tau_t-3\xi_x)g=0,
\\
u_xu_{xx}\colon\quad &\eta g_{uu}+(\eta_{uu}-3\xi_{ux})g+(\eta_u+\tau_t-3\xi_x)g_u=0,
\\
u_{xx}\colon\quad &(\eta_{ux}-\xi_{xx})g+\eta_xg_u=0,
\\
u_x^{\,3}\colon\quad &\eta g_{uuu}+(2\eta_u+\tau_t-3\xi_x)g_{uu}+3(\eta_{uu}-2\xi_{ux})g_u+(\eta_{uuu}-3\xi_{uux})g=0,
\\
u_x^{\,2}\colon\quad &\eta_x g_{uu}+(2\eta_{ux}-\xi_{xx})g_u+(\eta_{uux}-\xi_{uxx})g=0,
\\
u_x\colon\quad &\eta f_u+(\tau_t-\xi_x)f-\xi_t+(3\eta_{uxx}-\xi_{xxx})g+3\eta_{xx}g_u=0,
\\
1\colon\quad &\eta_t+\eta_x f+\eta_{xxx}g=0.
\end{align*}
Since $g\ne0$, the equations $(u_{xx}^{\,2}{:})$ and $(u_xu_{xxx}{:})$ give $\xi_u=0$,
and then the equations $(u_x^{\,2}u_{xx}{:})$ and~$(u_x^{\,4}{:})$ are identities.
The differential consequences $(u_xu_{xx}{:})-\p_u(u_{xxx}{:})$ and $(u_{xx}{:})-\p_x(u_{xxx}{:})$
take the form $\eta_{uu}g=0$ (i.e., $\eta_{uu}=0$) and $\eta_{xu}+2\xi_{xx}=0$.
Therefore,
\begin{gather}\label{eq:DeterEqComponents}
\tau=\tau(t),\quad \xi=\xi(t,x),\quad \eta=\eta^1u+\eta^0\quad\mbox{with}\quad\eta^1=-2\xi_x+\eta^{10},
\end{gather}
where  $\eta^0=\eta^0(t,x)$ and $\eta^{10}=\eta^{10}(t)$ are smooth functions of their arguments.
In view of the derived constraints for the components of~$Q$,
the equations $(u_x^{\,3}{:})$, $(u_x^{\,2}{:})$ and $(u_{xx}{:})$ coincide with the differential consequences $\p_u^2(u_{xxx}{:})$, $\p_u(u_{xx}{:})$ and $\p_x(u_{xxx}{:})$, respectively.
The equation $(u_x{:})$ can be replaced by its simplified version $(u_x{:})-3\p_x^2(u_{xxx}{:})$,
i.e., $\eta f_u+(\tau_t-\xi_x)f+2g\xi_{xxx}-\xi_t=0$.
As a result, the system of determining equations is equivalent to the representation~\eqref{eq:DeterEqComponents},
where the parameter functions satisfy the system of classifying equations
\begin{gather}\label{eq:SimplDeterEqI}
\bigl(\eta^1u+\eta^0\bigr)g_u+(\tau_t-3\xi_x)g=0,
\\\label{eq:SimplDeterEqII}
(\eta^1u+\eta^0)f_u+(\tau_t-\xi_x)f+2\xi_{xxx}g-\xi_t=0,
\\\label{eq:SimplDeterEqIII}
\bigl(\eta^1u+\eta^0\bigr)_t
+\bigl(\eta^1u+\eta^0\bigr)_xf
+\bigl(\eta^1u+\eta^0\bigr)_{xxx}g=0.
\end{gather}

\begin{proposition}\label{pro:Kernel}
The kernel Lie invariance algebra of the class~\eqref{eq:GenKdV} is $\mathfrak{g}^\cap=\langle\p_t,\p_x\rangle$.
\end{proposition}

\begin{proof}
Splitting the classifying equations~\eqref{eq:SimplDeterEqI}--\eqref{eq:SimplDeterEqIII}
with respect to the arbitrary elements~$f$ and~$g$ and their derivatives, which are then considered as independent variables,
we obtain $\eta^1=0$, $\eta^0=0$, $\tau_t=3\xi_x$, $\tau_t=\xi_x$, $\xi_t=0$ and $\xi_{xxx}=0$.
Hence $\eta=0$, $\tau_t=0$, $\xi_t=0$ and~$\xi_x=0$.
\end{proof}

\section{Group classification}\label{sec:GroupClassification}

Consider the subclass~$\mathcal L_{\rm HD}$ of the class~\eqref{eq:GenKdV} that is singled out by the constraints
\[
f=a+b|u+c|^{-3/2},\quad g=C|u+c|^{-3/2},
\]
where $a$, $b$, $c$ and~$C$ are arbitrary constants with $C\ne0$,
and denote its complement in the class~\eqref{eq:GenKdV} by~$\mathcal L_{\rm gen}$.
Both these subclasses are $G^\sim$-invariant.

\begin{theorem}\label{thm:GroupClassification}
A complete list of $G^\sim$-inequivalent Lie-symmetry extensions in the class~\eqref{eq:GenKdV}
is exhausted by the cases collected in Table~\ref{tab:GroupClassification}.
\end{theorem}

\begin{table}[!ht]
\begin{center}\small
\caption{The group classification of the class~\eqref{eq:GenKdV} up to the $G^\sim$-equivalence.
\strut}\label{tab:GroupClassification}
\renewcommand{\arraystretch}{1.5}
\begin{tabular}{|r|c|c|l|}
\hline
\hfil
no. &  $g$ &    $f$       &   \hfil Basis of $\mathfrak g_\theta$            \\
\hline
0 & $\forall$ & $\forall$ & $\p_t,\ \p_x$ \\
1 & $\forall$ & $0$ & $\p_t,\ \p_x,\ 3t\p_t+x\p_x$ \\
2 & $|u|^\mu$ & $\varepsilon |u|^\nu$ & $\p_t,\ \p_x,\ (\mu-3\nu)t\p_t+(\mu-\nu)x\p_x+2u\p_u$ \\
3 & $|u|^\mu$ & $\varepsilon\ln|u|$ & $\p_t,\ \p_x,\ \mu t\p_t+(\mu x+2\varepsilon t)\p_x+2u\p_u$ \\
4 & $e^{\mu u}$ & $\varepsilon e^u$ & $\p_t,\ \p_x,\ (\mu-3)t\p_t+(\mu-1)x\p_x+2\p_u$ \\
5 & $e^u$ & $\varepsilon u$ & $\p_t,\ \p_x,\ t\p_t+(x+2\varepsilon t)\p_x+2\p_u$ \\
6 & $|u|^\mu$ & $0$ & $\p_t,\ \p_x,\ 3t\p_t+x\p_x,\ \ \mu x\p_x+3u\p_u$ \\
7 & $e^u$ & $0$ & $\p_t,\ \p_x,\ 3t\p_t+x\p_x,\ \ x\p_x+3\p_u$ \\
8 & $1$ & $u$ & $\p_t,\ \p_x,\ 3t\p_t+x\p_x-2u\p_u,\ \ t\p_x+\p_u$ \\
9a & $|u|^{-3/2}$ & $0$ & $\p_t,\ \p_x,\ 3t\p_t+2u\p_u,\ x\p_x-2u\p_u,\ x^2\p_x-4xu\p_u$ \\
9b & $|u|^{-3/2}$ & $|u|^{-3/2}$ & $\p_t,\ \p_x,\ 3t\p_t+2u\p_u,\ \cos x\p_x+2\sin x\,u\p_u,\ \sin x\p_x-2\cos x\,u\p_u$ \\
9c & $|u|^{-3/2}$ & $-|u|^{-3/2}$ & $\p_t,\ \p_x,\ 3t\p_t+2u\p_u,\ \cosh x\p_x-2\sinh x\,u\p_u,\ \sinh x\p_x-2\cosh x\,u\p_u$ \\
\hline
\end{tabular}
\end{center}
\medskip
\parbox{\textwidth}{\small
Here $\varepsilon=\pm1$ mod~$G^\sim$, and $\mu$ and~$\nu$ are real constants.
In Case~0, $(f,g)$ is not $G^\sim$-equivalent to any other pair of functions from the table.
In Case~1, $g_u\ne0$, and $g$ is not $G^\sim$-equivalent to a power or an exponential function.
In Case~2, $\nu\ne0$ and $(\mu,\nu)\notin\{(0,1),(-\frac32,-\frac32)\}$.
In Case~6, $\mu\ne0,-\frac32$.
In Cases 2, 3, 6 and 9a–9c, the domain of $u$ does not contain zero unless $g=1$.
For $g=1$ (Case 2 with $\mu=0$), if this domain contains zero, then $\nu\in\mathbb N$ and $|u|^\nu$ is replaced by $u^\nu$, where $\varepsilon=1$ for odd $\nu$.
Cases 9a, 9b and~9c correspond to the equations~$\mathcal E_0$ (the Harry Dym equation), $\mathcal E_1$ and~$\mathcal E_{-1}$, respectively,
see~\eqref{eq:Eb}.
In each of these cases, the basis of~$\mathfrak g_\theta$ consists of $\p_t$, $3t\p_t+2u\p_u$ and the vector fields $\sigma\p_x-2\sigma_xu\p_u$,
where $\sigma$ runs through a fundamental set of solutions of the equation $\sigma_{xxx}+b\sigma_x=0$ with $b=0$, $b=1$ and $b=-1$, respectively.
}
\end{table}

\begin{proof}
We apply the method of furcate splitting, see the introduction for references.
In Section~\ref{sec:DetermEquat}, we have reduced the system of determining equations
for Lie symmetries of equations from the class~\eqref{eq:GenKdV} to the form
that is convenient for this method.
It consists of the representation~\eqref{eq:DeterEqComponents} for the components of~$Q$
and the classifying equations~\eqref{eq:SimplDeterEqI}--\eqref{eq:SimplDeterEqIII}.

The first step of the method is to derive the template form of systems for the arbitrary elements $(f,g)$.
For each Lie-symmetry vector field~$Q$ of the equation~$\mathcal L_\theta$ and each fixed value of $(t,x)$,
from the subsystem of the classifying equations~\eqref{eq:SimplDeterEqI} and~\eqref{eq:SimplDeterEqII}
we obtain a system with respect to the arbitrary-element tuple~$\theta=(f,g)$.
The \emph{template form} of such systems is
\begin{gather}\label{eq:TemplFormSimplDeterEqI}
(\alpha u+\beta)g_u=\mu g,
\\\label{eq:TemplFormSimplDeterEqII}
(\alpha u+\beta)f_u=\nu f+\delta g+\kappa,
\end{gather}
where the constant parameters $\alpha$, $\beta$, $\mu$, $\nu$, $\delta$ and~$\kappa$ are associated with
$\eta^1$, $\eta^0$, $3\xi_x-\tau_t$, $\xi_x-\tau_t$, $-2\xi_{xxx}$, $\xi_t$, respectively.
For a fixed value of the arbitrary-element tuple~$\theta$, we denote by~$V_\theta$ the linear span of the parameter tuples
$\vartheta:=(\alpha,\beta,\mu,\nu,\delta,\kappa)\in\mathbb R^6$
that are obtained from the classifying equations~\eqref{eq:SimplDeterEqI} and~\eqref{eq:SimplDeterEqII}
when $Q$ runs through~$\mathfrak g_\theta$ and $(t,x)$ runs through the domain of the variables.
Since the template form~\eqref{eq:TemplFormSimplDeterEqI}--\eqref{eq:TemplFormSimplDeterEqII} is linear in~$\vartheta$,
the tuple~$\theta$ satisfies the template-form system for each $\vartheta\in V_\theta$.
The number $k=k(\theta):=\dim V_\theta$ is the maximal number of template-form systems
with linearly independent tuples~$\vartheta$ that are given by Lie symmetries of~$\mathcal L_\theta$.
Transformations from~$G^\sim$ linearly change the parameters of template-form systems,
and thus the number~$k$ is $G^\sim$-invariant.

Consider different values of~$k$ separately.

\medskip\noindent
$\boldsymbol{k=0.}$
Then $\eta^1=\eta^0=0$, $3\xi_x-\tau_t=\xi_x-\tau_t=0$, $\xi_{xxx}=0$ and $\xi_t=0$ for any $Q\in\mathfrak g_\theta$,
and thus $\tau_t=0$ and $\xi_x=\xi_t=0$,
i.e., $\mathfrak g_\theta=\mathfrak g^\cap$, cf.\ the proof of Proposition~\ref{pro:Kernel}.
This is Case~0.

\medskip\noindent
$\boldsymbol{k=1.}$
Then $V_\theta=\langle\vartheta\rangle$ 
for a fixed nonzero tuple $\vartheta$,
and $\theta$ satisfies the corresponding template-form system~\eqref{eq:TemplFormSimplDeterEqI}--\eqref{eq:TemplFormSimplDeterEqII}.
For each $Q\in\mathfrak g_\theta$ and each $(t,x)$, we have
\[
\bigl(\eta^1,\eta^0,3\xi_x-\tau_t,\xi_x-\tau_t,-2\xi_{xxx},\xi_t\bigr)(t,x)=\Lambda(t,x)\vartheta
\]
for a function $\Lambda=\Lambda(t,x)$, and $\Lambda\not\equiv0$ if $Q\notin\mathfrak g^\cap$.
Since $\eta^1=-2\xi_x+\eta^{10}$, this means that
\begin{gather}\label{eq:k1Rels}
\begin{split}&
-2\xi_x+\eta^{10}=\Lambda\alpha,\quad
\eta^0=\Lambda\beta,\quad
3\xi_x-\tau_t=\Lambda\mu,\quad
\xi_x-\tau_t=\Lambda\nu,\\&
-2\xi_{xxx}=\Lambda\delta,\quad
\xi_t=\Lambda\kappa,
\end{split}
\end{gather}
and hence
\begin{gather}\label{eq:k1RelConseqs}
2\xi_x=\Lambda(\mu-\nu),\quad
2\tau_t=\Lambda(\mu-3\nu),\quad
\eta^{10}=\Lambda(\alpha+\mu-\nu).
\end{gather}

The further consideration splits into two cases depending on the values of $\mu-3\nu$ and $\alpha+\mu-\nu$.

\medskip\noindent
A.\ $\mu\ne3\nu$ or $\alpha+\mu-\nu\ne0$.
Then the second or the third equation of~\eqref{eq:k1RelConseqs} implies that $\Lambda=\Lambda(t)$,
and thus $\xi_x=\frac12\Lambda(\mu-\nu)$ depends only on~$t$, i.e., $\xi_{xx}=0$.
Hence $\Lambda\delta=-2\xi_{xxx}=0$, i.e., $\delta=0$, and
\[
\xi=\tfrac12\Lambda(\mu-\nu)x+\xi^0(t),\quad \xi^0_t=\Lambda\kappa,\quad \tau_t=\tfrac12\Lambda(\mu-3\nu),\quad \eta=\Lambda(\alpha u+\beta).
\]
The equation $\xi_{tx}=(\Lambda\kappa)_x=0$ gives $(\mu-\nu)\Lambda_t=0$.
The equation~\eqref{eq:SimplDeterEqIII} takes the form ${\Lambda_t(\alpha u+\beta)=0}$.
Therefore, $\Lambda_t=0$ if $(\alpha,\beta)\ne(0,0)$ or $\mu\ne\nu$.
We consider separately the subcases $\alpha\ne0$; $\alpha=0$, $\beta\ne0$; and $\alpha=\beta=0$.

\medskip\noindent
I.\ $\alpha\ne0$.
Rescaling the tuple $\vartheta$ 
and using shifts of~$u$ from~$G^\sim$, we set $\alpha=1$ and $\beta=0$.
Then $\eta^0=0$, $\eta=\Lambda u$ and hence $\Lambda_t=0$, i.e., $\Lambda=c_1$ is a constant.
The condition of Case~A means that $(\mu,\nu)\ne(-\frac32,-\frac12)$.
Integrating the template-form system $ug_u=\mu g$, $uf_u=\nu f+\kappa$, we obtain
\[
g=C_1|u|^\mu,\qquad f=C_2|u|^\nu-\frac\kappa\nu\ \ \mbox{if}\ \ \nu\ne0,\qquad f=\kappa\ln|u|+C_2\ \ \mbox{if}\ \ \nu=0,
\]
where $C_1\ne0$ and~$C_2$ are constants.
The constant summands in~$f$ are set to zero by Galilean boosts,
and the remaining constant multipliers are set to one in~$g$ and to $\varepsilon=\pm1$ in~$f$ by scalings and the reflection of~$t$ from~$G^\sim$.
If $\nu\ne0$ and $C_2\ne0$, we obtain Case~2.
If $\nu=0$ and $\kappa\ne0$, we obtain Case~3.
The corresponding extensions are given by
\[
\tau=\tfrac12c_1(\mu-3\nu)t+c_0,\quad \xi=\tfrac12c_1(\mu-\nu)x+c_1\kappa t+c_4,\quad \eta=c_1u,
\]
where $\kappa=0$ for Case~2 and $\kappa=\varepsilon$ for Case~3 after the gauge.
If $f$ is a constant, then $f=0$ up to the $G^\sim$-equivalence,
and $g=|u|^\mu$ with $\mu\ne0$ since the class~\eqref{eq:GenKdV} is nonlinear;
for such~$\theta$ we have $k\geqslant2$, see Cases~6 and~9a below.
For $(\mu,\nu)=(0,1)$, as well as for $\mu=\nu=-\frac32$, we also have $k\geqslant2$, see Cases~8, 9b and~9c below.

\medskip\noindent
II.\ $\alpha=0$, $\beta\ne0$.
Rescaling the tuple $\vartheta$, we set $\beta=1$.
Then $\eta^1=0$, $\eta=\Lambda$ and hence $\Lambda_t=0$, i.e., $\Lambda=c_1$ is a constant.
The template-form system $g_u=\mu g$, $f_u=\nu f+\kappa$ integrates to
\[
g=C_1e^{\mu u},\qquad f=C_2e^{\nu u}-\frac\kappa\nu\ \ \mbox{if}\ \ \nu\ne0,\qquad f=\kappa u+C_2\ \ \mbox{if}\ \ \nu=0.
\]
If $\nu\ne0$ and $C_2\ne0$, then we set $\nu=1$ by scaling~$u$, and up to the $G^\sim$-equivalence
we obtain $f=\varepsilon e^u$ and $g=e^{\mu u}$, where we re-denote $\mu/\nu$ by $\mu$ (Case~4).
If $\nu=0$, $\mu\ne0$ and $\kappa\ne0$, then we set $\mu=1$ by scaling~$u$,
which gives $f=\varepsilon u$ and $g=e^u$ up to the $G^\sim$-equivalence (Case~5).
The corresponding extensions are given by
\[
\tau=\tfrac12c_1(\mu-3\nu)t+c_0,\quad \xi=\tfrac12c_1(\mu-\nu)x+c_1\kappa t+c_4,\quad \eta=c_1,
\]
where $(\nu,\kappa)=(1,0)$ for Case~4 and $(\mu,\nu,\kappa)=(1,0,\varepsilon)$ for Case~5 after the gauge.
If $f$ is a constant, then $f=0$, $g=e^u$ up to the $G^\sim$-equivalence, and $k\geqslant2$ (Case~7 below).
If $\nu=\mu=0$, then $g$ is a constant and $f$ is linear in~$u$, which is the KdV equation with $k\geqslant2$ (Case~8 below).

\medskip\noindent
III.\ $\alpha=\beta=0$.
Then $\eta=0$, and the equation $\mu g=0$ implies $\mu=0$.
The condition of Case~A means that $\nu\ne0$, and thus $\Lambda_t=0$.
The template-form system takes the form $\nu f+\kappa=0$, i.e., $f$ is a constant,
and we set $f=0$ by a Galilean boost, which gives $\kappa=0$.
Hence $\tau=-\frac32c_1\nu t+c_0$, $\xi=-\frac12c_1\nu x+c_4$ and $\eta=0$,
and the arbitrary element~$g$ is not constrained, which gives Case~1.
If $g$ is $G^\sim$-equivalent to a power or an exponential function, then $k\geqslant2$, see Cases~6, 7 and~9a below.

\medskip\noindent
B.\ $\mu=3\nu$ and $\alpha+\mu-\nu=0$, i.e., $\mu=3\nu$ and $\alpha=-2\nu$.
Then the equations~\eqref{eq:k1Rels} and~\eqref{eq:k1RelConseqs} give $\tau_t=0$, $\eta^{10}=0$, $\xi_x=\Lambda\nu$ and $\eta=\Lambda(-2\nu u+\beta)$,
while $\Lambda$ may depend on~$x$.

\medskip\noindent
I.\ $\alpha\ne0$, i.e., $\nu\ne0$.
Rescaling the tuple $\vartheta$ and using shifts of~$u$, we set $\nu=-\frac12$ and $\beta=0$, and thus $\alpha=1$ and $\mu=-\frac32$.
Then $\eta^0=0$, $\eta=\Lambda u$ and $\xi_x=-\frac12\Lambda$.
The template-form system $ug_u=-\frac32g$, $uf_u=-\frac12f+\delta g+\kappa$ integrates to
\[
g=C_1|u|^{-3/2},\qquad f=C_2|u|^{-1/2}-\delta C_1|u|^{-3/2}+2\kappa,
\]
and the equation~\eqref{eq:SimplDeterEqIII} takes the form $\Lambda_t+f\Lambda_x+g\Lambda_{xxx}=0$.
Splitting it with respect to $|u|^{-1/2}$, $|u|^{-3/2}$ and~$1$, we derive
\[
C_2\Lambda_x=0,\qquad \Lambda_{xxx}-\delta\Lambda_x=0,\qquad \Lambda_t+2\kappa\Lambda_x=0.
\]
If $C_2\ne0$, then $\Lambda_x=0$ and $\Lambda_t=0$, i.e., $\Lambda=c_1$ is a constant, and thus $\xi_{xxx}=0$ and $\delta=0$.
Using equivalence transformations, we obtain Case~2 with $(\mu,\nu)=(-\frac32,-\frac12)$,
where $\tau=c_0$, $\xi=-\frac12c_1x+c_1\kappa t+c_4$ and $\eta=c_1u$ with $\kappa=0$ after the gauge.
If $C_2=0$, then $f=-\delta C_1|u|^{-3/2}+2\kappa$ and $g=C_1|u|^{-3/2}$, i.e., $\mathcal L_\theta\in\mathcal L_{\rm HD}$.
Using a Galilean boost, scalings and the reflection of~$t$ from~$G^\sim$, we set $\kappa=0$ and $C_1=1$
and reduce the equation~$\mathcal L_\theta$ to the equation
\begin{gather}\label{eq:Eb}
\mathcal E_b\colon\quad u_t+b|u|^{-3/2}u_x+\bigl(|u|^{-3/2}u_x\bigr)_{xx}=0
\end{gather}
with $b=-\delta$.
Then $\Lambda_t=0$, $\xi_t=\Lambda\kappa=0$, and the equation $-2\xi_{xxx}=\Lambda\delta$ with $\xi_x=-\frac12\Lambda$ gives
$\Lambda_{xx}+b\Lambda=0$.
Scaling~$x$ by a transformation from~$G^\sim$, we can set $b\in\{-1,0,1\}$.
Hence $\Lambda$ runs through the solution space of $\Lambda_{xx}+b\Lambda=0$, $\xi=-\frac12\int\Lambda\,{\rm d}x+c_2$ and $\eta=\Lambda u$,
which gives the vector fields $\p_x$ and $\sigma\p_x-2\sigma_xu\p_u$ with $\sigma_{xxx}+b\sigma_x=0$.
The equation~$\mathcal E_b$ also admits the vector field $3t\p_t+2u\p_u$,
whose tuple $(2,0,-3,-3,0,0)$ is not proportional to $\vartheta=(1,0,-\frac32,-\frac12,-b,0)$,
i.e., $k\geqslant2$ for~$\mathcal E_b$.
Solving the determining equations for~$\mathcal E_b$ directly (see also the case $k\geqslant2$ below),
we obtain Cases~9a, 9b and~9c for $b=0$, $b=1$ and $b=-1$, respectively.

\medskip\noindent
II.\ $\alpha=0$, i.e., $\nu=\mu=0$.
Then $\tau_t=0$, $\xi_x=0$, $\eta^1=0$, $\eta=\Lambda\beta$ and $\Lambda\delta=-2\xi_{xxx}=0$, i.e., $\delta=0$.
The template-form system is of the form $\beta g_u=0$ and $\beta f_u=\kappa$.
If $\beta=0$, then $\kappa=0$, which contradicts the condition that $\vartheta\ne0$. 
Hence $\beta\ne0$, and thus $g$ is a constant and $f$ is a linear function of~$u$ with $f_u\ne0$,
i.e., $\mathcal L_\theta$ is $G^\sim$-equivalent to the KdV equation, for which $k\geqslant2$ (Case~8 below).

\medskip\noindent
$\boldsymbol{k\geqslant2.}$
Consider the natural projection $\pi\colon V_\theta\to\mathbb R^2_{\alpha,\beta}$, $\vartheta\mapsto(\alpha,\beta)$.
If $\pi(\vartheta)=(0,0)$ for some $\vartheta\in V_\theta$, then the first equation of the corresponding template-form system is $\mu g=0$, i.e., $\mu=0$.
We consider two cases, $\dim\pi(V_\theta)=2$ and $\dim\pi(V_\theta)\leqslant1$.

\medskip\noindent
A.\ $\dim\pi(V_\theta)=2$.
Then $V_\theta$ contains tuples~$\vartheta_1$ and~$\vartheta_2$ with $(\alpha_1,\beta_1)=(1,0)$ and $(\alpha_2,\beta_2)=(0,1)$.
The first equations of the corresponding template-form systems, $ug_u=\mu_1g$ and $g_u=\mu_2g$,
imply $(\mu_2u-\mu_1)g=0$, i.e., $\mu_1=\mu_2=0$ and $g$ is a constant.
Then we set $g=1$ by a scaling from the equivalence group~$G^\sim$.
We also have $\mu=0$ for all tuples from~$V_\theta$, i.e., $\tau_t=3\xi_x$, which implies $\xi_{xx}=0$, and thus $\delta=0$ for any $\vartheta\in V_\theta$.
In particular, $\delta_1=\delta_2=0$.
Hence, the second equations of the template-form systems with~$\vartheta_1$ and~$\vartheta_2$ are $uf_u=\nu_1f+\kappa_1$ and $f_u=\nu_2f+\kappa_2$.
Substituting the expression for~$f_u$ in view of the second equation into the first one, we obtain $(\nu_2u-\nu_1)f=-\kappa_2u+\kappa_1$.
If $\nu_2\ne0$, then $f=-(\kappa_2u-\kappa_1)/(\nu_2u-\nu_1)$, $f_u=-(\kappa_1\nu_2-\kappa_2\nu_1)/(\nu_2u-\nu_1)^2$,
and the above second equation reduces to $(\kappa_1\nu_2-\kappa_2\nu_1)(\nu_2u-\nu_1+1)=0$.
Hence $\kappa_1\nu_2=\kappa_2\nu_1$, i.e., $f_u=0$, which contradicts the condition $f_u\ne0$ for $g=1$.
Therefore, $\nu_2=0$, and the equation $f_u=\kappa_2$ means that $f$ is a linear function of~$u$, i.e., $f=u$ up to the $G^\sim$-equivalence, which gives the KdV equation.
Then the equations~\eqref{eq:SimplDeterEqII} and~\eqref{eq:SimplDeterEqIII}
split with respect to~$u$ into the equations $\eta^1=-2\xi_x$, $\eta^0=\xi_t$, $\eta^1_t+\eta^0_x=0$ and $\eta^0_t+\eta^0_{xxx}=0$,
which implies $\tau_{tt}=0$, $\xi=\frac13\tau_tx+\xi^0(t)$ with $\xi^0_{tt}=0$.
This results in Case~8.

\medskip\noindent
B.\ $\dim\pi(V_\theta)\leqslant1$.
Since $k\geqslant2$, the kernel of~$\pi$ is nontrivial.
Hence, $V_\theta$ contains a nonzero tuple of the form $(0,0,0,\nu,\delta,\kappa)$, i.e., $\nu f+\delta g+\kappa=0$.
If $g$ is a constant, then $\mu=0$ for all tuples from~$V_\theta$, which implies $\delta=0$ as in Case~A for $k\geqslant2$,
and $\nu f+\kappa=0$ with $f_u\ne0$ gives $\nu=\kappa=0$, which is a contradiction.
Therefore, $g_u\ne0$ and then $\nu\ne0$.
Hence $f=a+bg$ for some constants~$a$ and~$b$, and we set $a=0$ by a Galilean boost, $f=bg$.
Since $\nu\ne0$ for each nonzero tuple from the kernel of~$\pi$, this kernel is one-dimensional, and thus $\dim\pi(V_\theta)=1$.
Hence, $V_\theta$ contains a tuple with $(\alpha,\beta)\ne(0,0)$, and $g$ satisfies the corresponding nontrivial equation $(\alpha u+\beta)g_u=\mu g$ with $\mu\ne0$.
Integrating it as in Case~A for $k=1$, we obtain, up to the $G^\sim$-equivalence, $g=|u|^\mu$ if $\alpha\ne0$ and $g=e^u$ if $\alpha=0$,
where the sign of~$g$ is set to be positive by the reflection of~$t$, which preserves the relation $f=bg$.
We solve the determining equations for these values of~$\theta$ directly.

If $g=|u|^\mu$ and $f=b|u|^\mu$,
then the equation~\eqref{eq:SimplDeterEqI} gives $\eta^0=0$ and $\mu\eta^1=3\xi_x-\tau_t$,
and differentiating the last equation with respect to~$x$, we get $(2\mu+3)\xi_{xx}=0$.
If $\mu\ne-\frac32$, then $\xi_{xx}=0$, and the equations~\eqref{eq:SimplDeterEqII} and~\eqref{eq:SimplDeterEqIII} give
$\eta^1_t=0$ and $\xi_t=0$.
If $b\ne0$, the equation~\eqref{eq:SimplDeterEqII} additionally gives $\mu\eta^1=\xi_x-\tau_t$, whence $\xi_x=0$ and $\tau_t=-\mu\eta^1$.
This results in Case~6 for $b=0$
and in Case~2 with $\nu=\mu$ for $b\ne0$, where we set $b=\varepsilon$ by scalings from~$G^\sim$.
If $\mu=-\frac32$, then $\mathcal L_\theta\in\mathcal L_{\rm HD}$, and the equation~$\mathcal L_\theta$ is reduced to the equation~\eqref{eq:Eb}.
The equation~\eqref{eq:SimplDeterEqI} gives $\eta^{10}=\frac23\tau_t$,
the equation~\eqref{eq:SimplDeterEqII} splits into $\xi_t=0$ and $\xi_{xxx}+b\xi_x=0$,
and then the equation~\eqref{eq:SimplDeterEqIII} implies $\tau_{tt}=0$.
Therefore, $\tau=c_1t+c_0$, $\eta=(\frac23c_1-2\xi_x)u$ and
$\xi=c_2+c_3x+c_4x^2$ for $b=0$,
$\xi=c_2+c_3\cos x+c_4\sin x$ for $b=1$ and $\xi=c_2+c_3\cosh x+c_4\sinh x$ for $b=-1$,
which gives Cases~9a, 9b and~9c, respectively.

If $g=e^u$ and $f=be^u$,
then the equation~\eqref{eq:SimplDeterEqI} gives $\eta^1=0$ and $\eta^0=3\xi_x-\tau_t$, whence $\xi_{xx}=0$,
and the equations~\eqref{eq:SimplDeterEqII} and~\eqref{eq:SimplDeterEqIII} give $\eta^0_t=0$ and $\xi_t=0$.
If $b\ne0$, the equation~\eqref{eq:SimplDeterEqII} additionally gives $\eta^0=\xi_x-\tau_t$, whence $\xi_x=0$ and $\tau_t=-\eta^0$.
This results in Case~7 for $b=0$
and in Case~4 with $\mu=1$ for $b\ne0$, where we set $b=\varepsilon$ by scalings from~$G^\sim$.

\medskip\noindent
Collecting the above results, we exclude the values of parameters for which the corresponding equations are considered in other cases.
In particular, the value $(\mu,\nu)=(-\frac32,-\frac32)$ is excluded from Case~2 and the value $\mu=-\frac32$ is excluded from Case~6
since the corresponding equations belong to~$\mathcal L_{\rm HD}$,
and the value $(\mu,\nu)=(0,1)$ is excluded from Case~2 since it corresponds to the KdV equation (Case~8).
Here we cannot set $\varepsilon=1$ in Cases 2--5 in general since $\tilde g=c_2^{\,3}g/c_1$ and $\tilde f=(c_2f+c_3)/c_1$,
and thus the transformations from~$G^\sim$ that preserve the condition $g>0$ satisfy $c_2/c_1>0$ and do not change the sign of~$f$ in these cases.
Finally, the listed cases are $G^\sim$-inequivalent since the functional form of the pairs $(f,g)$ and the number~$k$ are invariant with respect to transformations from~$G^\sim$,
and the values of the parameters
$\mu$, $\nu$ and~$\varepsilon$ in the canonical forms of Table~\ref{tab:GroupClassification} cannot be changed by these transformations.
\end{proof}

\begin{remark}\label{rem:RegularSingular}
Following~\cite{vane2020b}, we call the maximal Lie invariance algebra~$\mathfrak g_\theta$ of an equation~$\mathcal L_\theta$ from the class~\eqref{eq:GenKdV}
and the corresponding Lie-symmetry extension \emph{regular} in this class
if there exists a subalgebra~$\mathfrak s$ of~$\mathfrak g^\sim$ such that $\mathfrak g_\theta=\varpi_*\mathfrak s$, and \emph{singular} otherwise.
Here $\varpi$ denotes the natural projection from the space $\mathbb R^3_{t,x,u}\times\mathbb R^2_{f,g}$ onto the space $\mathbb R^3_{t,x,u}$.
The Lie-symmetry extensions from Cases~1--8 of Table~\ref{tab:GroupClassification} are regular.
At the same time, the Lie-symmetry extensions from Cases~9a--9c are singular
since the corresponding algebras~$\mathfrak g_\theta$ contain vector fields whose components are nonlinear in~$x$,
and such vector fields do not belong to~$\varpi_*\mathfrak g^\sim$.
In particular, all the Lie-symmetry extensions of the maximal dimension five are singular,
and hence they cannot be found by the algebraic method of group classification
based merely on the classification of subalgebras of~$\mathfrak g^\sim$, cf.~\cite{vane2020b}.
\end{remark}

\begin{remark}\label{rem:AlgebraTypes}
The structure of the algebras~$\mathfrak g_\theta$ presented in Table~\ref{tab:GroupClassification}
is determined by the commutation relations of the extension vector fields with $\p_t$ and~$\p_x$.
In the notation of~\cite{popo2003a}, the algebras~$\mathfrak g_\theta$ are of the following types:

\begin{center}{\renewcommand{\arraystretch}{1.25}%
\begin{tabular}[t]{@{}l@{\quad}l@{}}
$2A_1$ & Case~0\\
$A^{1/3}_{3.4}$ & Case~1\\
$A^a_{3.4}$ & Case~2$_{\mu\notin\{\nu,3\nu\}}$, \ Case~4$_{\mu\notin\{1,3\}}$\\
$A_{2.1}\oplus A_1$ & Case~2$_{\mu\in\{\nu,3\nu\}}$, \ Case~4$_{\mu\in\{1,3\}}$\\
$A_{3.1}$ & Case~3$_{\mu=0}$\\
$A_{3.2}$ & Case~3$_{\mu\ne0}$, \ Case~5
\end{tabular}\qquad\qquad
\begin{tabular}[t]{@{}l@{\quad}l@{}}
$2A_{2.1}$ & Cases~6 and~7\\
$A^{-2/3}_{4.8}$ & Case~8\\
$\mathrm{sl}(2,\mathbb R)\oplus A_{2.1}$ & Cases~9a--9c
\end{tabular}}
\end{center}
\noindent
Here $a=\alpha$ if $-1\leqslant\alpha<1$ and $a=\alpha^{-1}$ if $|\alpha|>1$
with $\alpha=\frac{\mu-\nu}{\mu-3\nu}$ in Case~2 and $\alpha=\frac{\mu-1}{\mu-3}$ in Case~4.
Note that $\alpha\ne1$ under conditions imposed on~$\mu$ and $\nu$.
\end{remark}

\begin{remark}\label{rem:FurcateSplitting}
Since $\Lambda\delta=-2\xi_{xxx}$, the parameter~$\delta$ of the template form~\eqref{eq:TemplFormSimplDeterEqII}
may be nonzero only if $\xi_{xxx}\ne0$.
In view of the relation $2\xi_x=\Lambda(\mu-\nu)$, this requires $\Lambda_x\ne0$,
which is possible only in Case~B with $k=1$, where $\mu=3\nu$ and $\alpha=-2\nu$.
In Case~B.II, we have $\xi_x=0$.
In Case~B.I, the splitting of the classifying equation~\eqref{eq:SimplDeterEqIII} with respect to $|u|^{-1/2}$, $|u|^{-3/2}$ and~$1$
gives $C_2\Lambda_x=0$, and thus $\Lambda_x\ne0$ only if $C_2=0$, i.e., only if $\mathcal L_\theta\in\mathcal L_{\rm HD}$.
Similarly, in the case $k\geqslant2$, the equality $\xi_{xx}=0$ fails only for the equations from~$\mathcal L_{\rm HD}$.
Therefore, in all the cases of the proof except those leading to the subclass~$\mathcal L_{\rm HD}$, we have $\xi_{xx}=0$, which implies $\delta=0$,
and the template-form equation for~$f$ reduces to $(\alpha u+\beta)f_u=\nu f+\kappa$.
In other words, the summand~$\delta g$ in the template form~\eqref{eq:TemplFormSimplDeterEqII} is needed only for singling out
the singular Lie-symmetry extensions, cf.\ Remark~\ref{rem:RegularSingular}.
\end{remark}

\begin{remark}
In Cases~2--9c, we gauge the sign of the coefficient of~$g$ to be positive
using the reflection $(t,f,g)\mapsto(-t,-f,-g)$ from~$G^\sim$.
After this gauge, the sign of~$f$ in Cases~2--5 cannot be changed by transformations from~$G^\sim$,
see the end of the proof of Theorem~\ref{thm:GroupClassification}, and therefore the parameter $\varepsilon=\pm1$ appears in~$f$.
In the same way, the equations~$\mathcal E_1$ and~$\mathcal E_{-1}$ from Cases~9b and~9c differ by the sign of~$f$.
\end{remark}

\begin{remark}
The subclass of the class~\eqref{eq:GenKdV} with $g=1$ consists of the generalized KdV equations $u_t+f(u)u_x+u_{xxx}=0$,
and the Lie-symmetry extensions in this subclass are given by Cases~2, 3 and~4 with $\mu=0$ and Case~8.
This subclass is mapped by the reflection $t\mapsto-t$ to a subclass of the class of KdV-type equations $u_t=u_{xxx}+F(t,x,u,u_x,u_{xx})$,
whose Lie symmetries were studied in~\cite{gung2004a}.
\end{remark}

\begin{remark}\label{rem:KmnClassification}
Consider the nonlinear Rosenau--Hyman equations $K(m,n)$ with $mn\ne0$ and $(m,n)\ne(1,1)$,
for which $f=mu^{m-1}$ and $g=nu^{n-1}$.
If $m\ne1$, then the equation $K(m,n)$ is $G^\sim$-equivalent to an equation from Case~2 with $\mu=n-1$ and $\nu=m-1$,
and the corresponding extension is $(n-3m+2)t\p_t+(n-m)x\p_x+2u\p_u$,
except for the equations $K(2,1)$ and $K(-\frac12,-\frac12)$.
The equation $K(2,1)$ is the KdV equation (Case~8),
and the equation $K(-\frac12,-\frac12)$ is $G^\sim$-equivalent to the equation~$\mathcal E_1$ (Case~9b).
If $m=1$, then the summand $u_x$ is removed by a Galilean boost,
and the equation $K(1,n)$ is $G^\sim$-equivalent to an equation from Case~6 with $\mu=n-1$ for $n\ne-\frac12$
and to the Harry Dym equation~$\mathcal E_0$ (Case~9a) for $n=-\frac12$.
The defocusing version $u_t-(u^{-1/2})_x+(u^{-1/2})_{xxx}=0$ of the equation $K(-\frac12,-\frac12)$ is $G^\sim$-equivalent to the equation~$\mathcal E_{-1}$ (Case~9c).
The integrability of the equation $K(-\frac12,-\frac12)$ was known, see~\cite{rose1997a,vodo2013a}.
In~\cite{rose1997a}, this equation was related to the equation $K(\frac32,-\frac12)$ by the Lagrange map,
and the equations $K(n+2,n)$ were related to generalized Harry Dym equations.
In~\cite{hern2020a}, the equations $u_t+(u^m)_x+(u^{-1/2})_{xxx}=0$ with $m\in\{-\frac12,0,1,\frac32\}$,
including the equation $K(-\frac12,-\frac12)$ and the Harry Dym equation,
were mapped by a chain of potentiation, a hodograph transformation, antipotentiation and the point transformation $u\mapsto e^u$
to particular cases of the Calogero--Degasperis--Fokas equation $v_t=v_{xxx}-\frac12v_x^{\,3}+(\alpha e^{2v}+\beta e^{-2v}+\gamma)v_x$,
which is in turn related to the KdV equation by a Miura-type transformation.
For the Harry Dym equation, similar connections with the modified KdV and the KdV equations were found in~\cite{here1989a,kawa1985a}.
All these transformations are nonlocal.
\end{remark}

\begin{theorem}\label{thm:GroupClassificationGroupoid}
Up to the point equivalence, cases of Lie-symmetry extensions in the class~\eqref{eq:GenKdV}
are exhausted by Cases 1--8 and~9a of Table~\ref{tab:GroupClassification}.
More specifically, the only additional point equivalence transformations between equations from different cases of Table~\ref{tab:GroupClassification}
are those between Cases~9a, 9b and~9c.
Namely, the equations from Cases~9b and~9c are mapped to the Harry Dym equation (Case~9a)
by the following point transformations:
\begin{gather}\label{eq:HDMappingB}
\mbox{9b}\to\mbox{9a}\colon\ \hat t=t,\ \ \hat x=2\tan\frac x2,\ \ \hat u=\cos^4\!\frac x2\,u,
\\\label{eq:HDMappingC}
\mbox{9c}\to\mbox{9a}\colon\ \hat t=t,\ \ \hat x=2\tanh\frac x2,\ \ \hat u=\cosh^4\!\frac x2\,u
\quad\mbox{or}\quad \hat t=t,\ \ \hat x=e^x,\ \ \hat u=e^{-2x}u.
\end{gather}
\end{theorem}

\begin{proof}
It is checked by direct computation that the transformations~\eqref{eq:HDMappingB} and~\eqref{eq:HDMappingC} (locally) map
the equations from Cases~9b and~9c to the Harry Dym equation from Case~9a, respectively.
Therefore, it suffices to show that equations from different cases of Table~\ref{tab:GroupClassification}, except Cases~9a--9c,
are not related by point transformations.

Suppose that a point transformation~$\Phi$ maps an equation~$\mathcal L_\theta$ to an equation~$\mathcal L_{\tilde\theta}$,
where the arbitrary-element tuples~$\theta$ and~$\tilde\theta$ are taken from Table~\ref{tab:GroupClassification}.
Since the pushforward~$\Phi_*$ by~$\Phi$ is an isomorphism of the algebras~$\mathfrak g_\theta$ and~$\mathfrak g_{\tilde\theta}$,
the dimensions and the types of these algebras presented in Remark~\ref{rem:AlgebraTypes}
are invariant with respect to point transformations.
The algebras~$\mathfrak g_\theta$ are isomorphic only for equations from Cases~1, 2 and~4,
from Cases~3$_{\mu\ne0}$ and~5 and from Cases~6 and~7.

As shown in the proof of Theorem~\ref{thm:EquivGroup},
the transformation~$\Phi$ is of the form $\tilde t=T(t)$, $\tilde x=X(t,x)$, $\tilde u=U^1(t,x)u+U^0(t,x)$,
and its components satisfy the equations~\eqref{eq:AdmTransDetEqA} and \eqref{eq:AdmTransDetEqCReduced}--\eqref{eq:AdmTransDetEqE}.
The sum of the derived algebra and the center of~$\mathfrak g_\theta$ is a megaideal of~$\mathfrak g_\theta$.
It is equal to $\langle\p_t,\p_x\rangle$ for each of Cases~1--7 except Case~3$_{\mu=0}$ and to $\langle\p_x\rangle$ in Case~3$_{\mu=0}$.
The pushforward~$\Phi_*$ of vector fields by~$\Phi$ establishes an isomorphism between~$\mathfrak g_\theta$ and~$\mathfrak g_{\tilde\theta}$
and thus maps the above megaideal onto the corresponding megaideal of~$\mathfrak g_{\tilde\theta}$.
This is why we have $\Phi_*\langle\p_t,\p_x\rangle=\langle\p_{\tilde t},\p_{\tilde x}\rangle$ for Cases~1--7 except Case~3$_{\mu=0}$,
and $\Phi_*\langle\p_x\rangle=\langle\p_{\tilde x}\rangle$ for Case~3$_{\mu=0}$.
Here
\[
\Phi_*\p_t=T_t\p_{\tilde t}+X_t\p_{\tilde x}+U_t\p_{\tilde u},\quad
\Phi_*\p_x=X_x\p_{\tilde x}+U_x\p_{\tilde u}.
\]
Therefore, $X_x$ is a constant and $U_x=0$ in all the cases.
Except Case~3$_{\mu=0}$, the vector field $\Phi_*\p_t$ is a linear combination of~$\p_{\tilde t}$ and~$\p_{\tilde x}$ with constant coefficients,
i.e., $T_t$ and~$X_t$ are constants and $U_t=0$.
In Case~3$_{\mu=0}$, the equation~\eqref{eq:AdmTransDetEqE} with $U_x=0$ gives $U_t=0$,
the equations~\eqref{eq:AdmTransDetEqD} and~\eqref{eq:AdmTransDetEqE} reduce to~\eqref{eq:AdmTransDetEqCaseA},
and, in the same way as in the proof of Theorem~\ref{thm:EquivGroup}, we derive $T_{tt}=X_{tt}=0$.
As a result, in all the cases the transformation~$\Phi$ is of the form~\eqref{eq:EquivGroupTransformA}
and is thus the projection of an element of~$G^\sim$.
Hence the tuples~$\theta$ and~$\tilde\theta$ are $G^\sim$-equivalent,
i.e., they belong to the same case of Table~\ref{tab:GroupClassification} with the same values of the parameters $\mu$, $\nu$ and~$\varepsilon$,
see the end of the proof of Theorem~\ref{thm:GroupClassification}.
\end{proof}

\begin{remark}\label{rem:HDMappingsExpected}
The existence of the transformations~\eqref{eq:HDMappingB} and~\eqref{eq:HDMappingC} is expected and in agreement with the structure of the corresponding maximal Lie invariance algebras.
The algebras from Cases~9a, 9b and~9c are isomorphic,
and the pushforwards of the algebras~$\mathfrak g_{\mathcal E_1}$ and~$\mathfrak g_{\mathcal E_{-1}}$
by the transformations~\eqref{eq:HDMappingB} and~\eqref{eq:HDMappingC}, respectively, coincide with~$\mathfrak g_{\mathcal E_0}$.
For example, the transformation $\hat x=2\tan\frac x2$, $\hat u=\cos^4\frac x2\,u$ maps the vector fields
$\p_x$, $\cos x\,\p_x+2\sin x\,u\p_u$ and $\sin x\,\p_x-2\cos x\,u\p_u$ from Case~9b
to $(1+\frac14\hat x^2)\p_{\hat x}-\hat x\hat u\p_{\hat u}$, $(1-\frac14\hat x^2)\p_{\hat x}+\hat x\hat u\p_{\hat u}$ and $\hat x\p_{\hat x}-2\hat u\p_{\hat u}$, respectively,
which span the same algebra as $\p_{\hat x}$, $\hat x\p_{\hat x}-2\hat u\p_{\hat u}$ and $\hat x^2\p_{\hat x}-4\hat x\hat u\p_{\hat u}$.
At the same time, these algebras are realized in different ways, which are inequivalent with respect to~$G^\sim$:
components of vector fields from the algebras~$\mathfrak g_{\mathcal E_0}$, $\mathfrak g_{\mathcal E_1}$ and~$\mathfrak g_{\mathcal E_{-1}}$
involve power, trigonometric and hyperbolic functions of~$x$, respectively.
This is a manifestation of the fact that the transformations~\eqref{eq:HDMappingB} and~\eqref{eq:HDMappingC} are not induced by elements of~$G^\sim$.
Their existence  can be justified by deeper arguments.
For each $b\in\{-1,0,1\}$, the vector fields $\sigma\p_x-2\sigma_xu\p_u$ with $\sigma_{xxx}+b\sigma_x=0$ span a subalgebra of~$\mathfrak g_{\mathcal E_b}$ isomorphic to ${\rm sl}(2,\mathbb R)$.
The projections of these subalgebras to the $x$-line,
\[
\langle\p_x,x\p_x,x^2\p_x\rangle,\quad \langle\p_x,\cos x\,\p_x,\sin x\,\p_x\rangle\quad\mbox{and}\quad \langle\p_x,\cosh x\,\p_x,\sinh x\,\p_x\rangle,
\]
are locally equivalent realizations of ${\rm sl}(2,\mathbb R)$ in view of Lie's classification of realizations of Lie algebras on the line, see, e.g.,~\cite{boyk2021a}.
More specifically, the point transformation $\hat t=t$, $\hat x=X(x)$, $\hat u=X_x^{\,-2}u$ with $X_x>0$ maps the equation~$\mathcal E_b$ to the equation~$\mathcal E_0$
if the Schwarzian derivative $X_{xxx}/X_x-\frac32(X_{xx}/X_x)^2$ of~$X$ is equal to~$b/2$.
This condition defines the function~$X$ up to orientation-preserving linear fractional transformations of~$\hat x$,
which are combined with the corresponding transformations of~$\hat u$ to point symmetries of the Harry Dym equation~$\mathcal E_0$.
In particular, the two forms of the transformation~\eqref{eq:HDMappingC} are related in this way, $2\tanh\frac x2=2(e^x-1)/(e^x+1)$.
We emphasize that the transformations~\eqref{eq:HDMappingB} and~\eqref{eq:HDMappingC} are local (e.g., the transformation~\eqref{eq:HDMappingB} is defined only for $|x|<\pi$)
and, in particular, do not preserve periodicity conditions.
\end{remark}

\begin{remark}\label{rem:KmnHD}
In view of Theorem~\ref{thm:GroupClassificationGroupoid} and Remark~\ref{rem:KmnClassification},
the equation $K(-\frac12,-\frac12)$ and its defocusing counterpart are similar to the Harry Dym equation
with respect to the compositions of transformations from~$G^\sim$ with the point transformations~\eqref{eq:HDMappingB} and~\eqref{eq:HDMappingC},
respectively.
Up to scalings, the defocusing version coincides with the extended Harry Dym equation $v_t=-\frac12(\p_x^2-\kappa^2)(v^{-1/2})_x$, which was related to the Harry Dym equation by the second form of the point transformation~\eqref{eq:HDMappingC} in~\cite{marv2017a}, where the multi-component generalization was also considered.
To the best of our knowledge, the point transformation~\eqref{eq:HDMappingB}, which relates the equation $K(-\frac12,-\frac12)$ itself to the Harry Dym equation, has not been presented in the literature.
\end{remark}

\begin{remark}\label{rem:GeometricMeaning}
The transformations from the one-parameter groups generated by the vector fields given in Table~\ref{tab:GroupClassification}
have a simple geometric meaning, cf.~\cite[Example~2.41]{olve1993A}.
They are independent shifts of~$t$ and~$x$,
the Galilean boosts in Case~8,
various scalings of~$(t,x)$ combined with either scalings or shifts of~$u$
and, in Cases~3 and~5, additionally composed with Galilean boosts.
In Case 9a, independent shifts and scalings of $t$ and $x$ with appropriate scalings of $u$ are supplemented by the projective transformations $\tilde t=t$, $\tilde x=x/(1-\epsilon x)$, $\tilde u=(1-\epsilon x)^4u$ with an arbitrary constant~$\epsilon$, which are conjugate to shifts of $x$ by the discrete point symmetry $\tilde t=t$, $\tilde x=-x^{-1}$, $\tilde u=x^4u$ of the Harry Dym equation.
In Cases~9b and~9c, the one-parameter Lie symmetry groups are conjugate to those in Case~9a
by the transformations~\eqref{eq:HDMappingB} and~\eqref{eq:HDMappingC}, respectively.
\end{remark}

\section{Conservation laws of nonpositive order}\label{sec:CLs}

For a theoretical background on conservation laws and their characteristics, see~\cite{blum2009A,olve1993A}.
Each equation from the class~\eqref{eq:GenKdV} is in conserved form and thus admits a conservation law with a conserved density of nonpositive order.
Evolution equations in two independent variables that admit one or two low-order conservation laws can be reduced to certain normal forms,
see~\cite{popo2010c}, where, in particular, the Harry Dym equation and Korteweg--de Vries-type equations were considered.
Each equation~$\mathcal L_\theta$ from the class~\eqref{eq:GenKdV} can be represented in the Kovalevska form $u_t=-fu_x-(gu_x)_{xx}$.
Therefore, we consider the derivatives of~$u$ involving differentiation with respect to~$t$ and the other derivatives of~$u$
as the principal and the parametric derivatives of~$\mathcal L_\theta$, respectively.
A conserved current of~$\mathcal L_\theta$ is a tuple $(\rho,\sigma)$ of differential functions of~$u$
such that ${\rm D}_t\rho+{\rm D}_x\sigma=0$ on the solution set of~$\mathcal L_\theta$.
Up to the equivalence of characteristics, i.e., modulo adding trivial characteristics, which vanish on the solutions of~$\mathcal L_\theta$,
we can consider conservation-law characteristics of~$\mathcal L_\theta$ to be differential functions~$\lambda\{u\}$,
where the notation $\lambda\{u\}$ means that $\lambda$ depends only on $(t,x)$ and parametric derivatives of~$\mathcal L_\theta$,
and such a characteristic is the canonical representative of its equivalence class.
A differential function~$\lambda\{u\}$ is a conservation-law characteristic of~$\mathcal L_\theta$ if and only if
the product $\lambda\bigl(u_t+fu_x+(gu_x)_{xx}\bigr)$ is a total divergence,
$\lambda\bigl(u_t+fu_x+(gu_x)_{xx}\bigr)={\rm D}_t\rho+{\rm D}_x\sigma$ for some differential functions~$\rho$ and~$\sigma$ of~$u$,
i.e., if $\mathsf E\bigl(\lambda(u_t+fu_x+(gu_x)_{xx})\bigr)=0$~\cite{anco2002b,olve1993A}.
Here $\mathsf E$ denotes the Euler operator.
Then the tuple $(\rho,\sigma)$ is a conserved current of~$\mathcal L_\theta$
that is contained in the conservation law with the characteristic~$\lambda$.
Since the equation~$\mathcal L_\theta$ is a normal, totally nondegenerate system of differential equations,
in view of~\cite[Theorem~4.26]{olve1993A}, a conserved current of~$\mathcal L_\theta$ is trivial if and only if the associated characteristic identically vanishes,
and thus the space of conservation laws of~$\mathcal L_\theta$ is naturally isomorphic to the space of canonical representatives of its conservation-law characteristics.
The order~${\rm ord}\,\lambda$ of a differential function~$\lambda$ is the highest order of derivatives of~$u$ involved in~$\lambda$,
and ${\rm ord}\,\lambda:=-\infty$ if $\lambda$ does not depend on~$u$.
We classify the conservation laws of equations from the class~\eqref{eq:GenKdV} whose characteristics are of nonpositive order,
i.e., they depend at most on $(t,x,u)$.
These are the simplest conservation laws, and their classification is much simpler than the group classification.
At the same time, the integrable equations from the class~\eqref{eq:GenKdV}, like the KdV, the modified KdV and the Harry Dym equations,
admit infinite series of local conservation laws of increasing orders, see, e.g.,~\cite{olve1993A},
and their description is beyond the scope of the present paper.

\begin{theorem}\label{thm:CLs}
A complete list of $G^\sim$-inequivalent cases for the space $\Lambda_\theta$ of canonical representatives of equivalence classes
of conservation-law characteristics of nonpositive order of equations~$\mathcal L_\theta$ from the class~\eqref{eq:GenKdV}
is exhausted by the cases collected in Table~\ref{tab:CLChars}.
\end{theorem}

\begin{table}[!ht]
\begin{center}\small
\caption{Classification of nonpositive-order conservation-law characteristics in the class~\eqref{eq:GenKdV} up to the $G^\sim$-equivalence.
\strut}\label{tab:CLChars}
\renewcommand{\arraystretch}{1.4}
\begin{tabular}{|r|c|c|l|}
\hline
no. & $g$ & $f$ & \hfil Basis of $\Lambda_\theta$\\
\hline
0 & $\forall$ & $\forall$ & $1,\ \int g{\rm d}u$\\
1 & $\forall$ & $0$ & $1,\ \int g{\rm d}u,\ x,\ x^2$\\
2 & $\forall$ & $g$ & $1,\ \int g{\rm d}u,\ \cos x,\ \sin x$\\
3 & $\forall$ & $-g$ & $1,\ \int g{\rm d}u,\ \cosh x,\ \sinh x$\\
4 & $\forall$ & $\int g{\rm d}u$ & $1,\ \int g{\rm d}u,\ x-t\int g{\rm d}u$\\
5 & $e^u$ & $0$ & $1,\ e^u,\ x,\ x^2,\ te^u-\frac16x^3$\\
6 & $e^u$ & $e^u$ & $1,\ e^u,\ \cos x,\ \sin x,\ x-te^u$\\
7 & $e^u$ & $-e^u$ & $1,\ e^u,\ \cosh x,\ \sinh x,\ x+te^u$\\
\hline
\end{tabular}
\end{center}
\medskip
\parbox{\textwidth}{\small
Here $\int g{\rm d}u$ is a fixed antiderivative of~$g$.
In Case~0, $(f,g)$ is $G^\sim$-inequivalent to any other pair of functions from the table.
In Cases~1--3, $g\ne\const$ and $g\notin\langle e^u\rangle$, and in Case~4, $g$ is $G^\sim$-inequivalent to~$e^u$.
}
\end{table}

\begin{proof}
Applying the Euler operator~$\mathsf E$ to $\lambda(t,x,u)\bigl(u_t+fu_x+(gu_x)_{xx}\bigr)$ and splitting it with respect to the derivatives of~$u$,
we obtain the system $\lambda_{ux}=0$, $g\lambda_{uu}=g_u\lambda_u$ and $\lambda_t+f\lambda_x+g\lambda_{xxx}=0$,
the other equations being its differential consequences.
Hence $\lambda=\chi(t)\int g{\rm d}u+\lambda^0(t,x)$, and the last equation takes the form
\begin{gather}\label{eq:CLClassifyingEq}
\chi_t\int g{\rm d}u+\lambda^0_t+f\lambda^0_x+g\lambda^0_{xxx}=0.
\end{gather}
For each fixed $(t,x)$, this is a linear relation between the functions $1$, $f$, $g$ and~$\int g{\rm d}u$ of~$u$
with the coefficients $\lambda^0_t$, $\lambda^0_x$, $\lambda^0_{xxx}$ and~$\chi_t$.
The functions~$1$ and~$\int g{\rm d}u$ are linearly independent since $g\ne0$.
We consider two cases depending on whether or not the functions $1$, $g$ and~$\int g{\rm d}u$ are linearly independent.

Suppose that $1$, $g$ and~$\int g{\rm d}u$ are linearly independent.
If $f\notin\langle1,g,\int g{\rm d}u\rangle$, then the functions $1$, $f$, $g$ and~$\int g{\rm d}u$ are linearly independent,
and the equation~\eqref{eq:CLClassifyingEq} implies $\chi_t=\lambda^0_t=\lambda^0_x=\lambda^0_{xxx}=0$,
i.e., $\Lambda_\theta=\langle1,\int g{\rm d}u\rangle$, which is Case~0 of Table~\ref{tab:CLChars}.
Otherwise, $f=\beta_0+\beta_2g+\beta_3\int g{\rm d}u$ for some constants $\beta_0$, $\beta_2$ and~$\beta_3$,
where $\beta_0$ can be set to zero by a Galilean boost from~$G^\sim$,
and the equation~\eqref{eq:CLClassifyingEq} splits into $\lambda^0_t=0$, $\lambda^0_{xxx}+\beta_2\lambda^0_x=0$ and $\chi_t+\beta_3\lambda^0_x=0$.
If $\beta_3\ne0$, then $\lambda^0_x$ is a constant,
and hence $\lambda^0$ is a linear function of~$x$ with $\beta_2\lambda^0_x=0$.
If $\beta_2\ne0$, then $\lambda^0_x=0$ and $\Lambda_\theta=\langle1,\int g{\rm d}u\rangle$.
If $\beta_2=0$, then $\Lambda_\theta=\langle1,\int g{\rm d}u,x-\beta_3t\int g{\rm d}u\rangle$, and we can set $\beta_3=1$ by scalings from~$G^\sim$,
since these scalings multiply~$\beta_3$ by $(c_5c_2^{\,2})^{-1}$.
This gives Case~4 of Table~\ref{tab:CLChars}.
If $\beta_3=0$, then $\chi$ is a constant and $\lambda^0=\lambda^0(x)$ is an arbitrary solution of $\lambda^0_{xxx}+\beta_2\lambda^0_x=0$.
Scalings of~$x$ from~$G^\sim$ multiply~$\beta_2$ by $c_2^{\,-2}$, and thus we can set $\beta_2\in\{0,1,-1\}$,
which gives Cases~1, 2 and~3 of Table~\ref{tab:CLChars} for $\beta_2=0$, $\beta_2=1$ and $\beta_2=-1$.

Suppose that $1$, $g$ and~$\int g{\rm d}u$ are linearly dependent.
Then up to the $G^\sim$-equivalence, $g=1$, $\int g{\rm d}u=u$ or $g=e^u$, $\int g{\rm d}u=e^u$.

For $g=1$, we have $f_u\ne0$ since the class~\eqref{eq:GenKdV} is nonlinear,
and the equation~\eqref{eq:CLClassifyingEq} takes the form $\chi_tu+\lambda^0_t+f\lambda^0_x=0$.
If $f\notin\langle1,u\rangle$, then $\chi_t=\lambda^0_t=\lambda^0_x=0$, i.e., $\Lambda_\theta=\langle1,u\rangle$ as in Case~0.
Otherwise, $f=\beta u$ with $\beta\ne0$ after a Galilean boost and we set $\beta=1$ by scalings from~$G^\sim$.
This is Case~4 of Table~\ref{tab:CLChars} for $g=1$, i.e., the KdV equation.

For $g=e^u$, the equation~\eqref{eq:CLClassifyingEq} takes the form $\lambda^0_t+f\lambda^0_x+e^u(\chi_t+\lambda^0_{xxx})=0$.
If $f\notin\langle1,e^u\rangle$, then $\lambda^0_x=0$, and thus $\lambda^0_t=\chi_t=0$, i.e., $\Lambda_\theta=\langle1,e^u\rangle$ as in Case~0.
Otherwise, setting the constant summand of~$f$ to zero, we obtain $f=\beta e^u$ and then $\lambda^0_t=0$ and $\lambda^0_{xxx}+\beta\lambda^0_x=-\chi_t$.
Hence $\chi_t$ is a constant, $\chi_t=k$.
The transformations from~$G^\sim$ that preserve the equality $g=e^u$ multiply~$\beta$ by $c_2^{\,-2}$,
and thus we can set $\beta\in\{0,1,-1\}$.
For $\beta=0$, the additional characteristics are $x$, $x^2$ and $te^u-\frac16x^3$ (for $k=1$), which is Case~5 of Table~\ref{tab:CLChars}.
For $\beta=\pm1$, the additional characteristics are the solutions of $\lambda^0_{xxx}\pm\lambda^0_x=0$ that differ from constants
and $te^u-\beta^{-1}x$ (for $k=1$), which gives Cases~6 and~7 of Table~\ref{tab:CLChars}.
\end{proof}

For each of the characteristics from Theorem~\ref{thm:CLs}, a conserved current $(\rho,\sigma)$
that is contained in the conservation law with this characteristic is presented in Table~\ref{tab:CLs}.

\begin{table}[!ht]
\begin{center}\small
\caption{Conserved currents associated with nonpositive-order characteristics for equations from the class~\eqref{eq:GenKdV}.\strut}\label{tab:CLs}
\renewcommand{\arraystretch}{1.4}
\begin{tabular}{|l|l|l|l|}
\hline
\hfil Constraint & $\hfil \lambda$ & $\hfil \rho$ & $\hfil \sigma$\\
\hline
--- & $1$ & $u$ & $(gu_x)_x+\int f{\rm d}u$\\[1ex]
--- & $\int g{\rm d}u$ & $\iint g{\rm d}u{\rm d}u$ & $(gu_x)_x\int g{\rm d}u-\frac12g^2u_x^{\,2}+\int\bigl(f\int g{\rm d}u\bigr){\rm d}u$\\[1ex]
$f=\beta g$ & $s$ & $su$ & $s(gu_x)_x-s_xgu_x+(s_{xx}+\beta s)\int g{\rm d}u$\\[1ex]
$f=\int g{\rm d}u$ & $x-t\int g{\rm d}u$ & $xu-t\iint g{\rm d}u{\rm d}u$ & $\big(x-t\int g{\rm d}u\big)(gu_x)_x+\frac12tg^2u_x^{\,2}-gu_x$\\
& & & ${}+x\iint g{\rm d}u{\rm d}u-t\int\bigl(\int g{\rm d}u\bigr)^2{\rm d}u$\\[1ex]
$f=0$, $g=e^u$ & $te^u-\frac16x^3$ & $te^u-\frac16x^3u$ & $\big(te^u-\frac16x^3\big)(gu_x)_x-\frac12 tg^2u_x^{\,2}+\frac12x^2gu_x-xe^u$\\[1ex]
\hline
\end{tabular}
\end{center}
\medskip
\parbox{\textwidth}{\small
Here $s=s(x)$ is an arbitrary solution of the equation $s_{xxx}+\beta s_x=0$.}
\end{table}

\begin{remark}
For the Harry Dym equation~$\mathcal E_0$ and, more generally, for the equations~$\mathcal E_b$,
the space $\Lambda_\theta$ is four-dimensional,
which is in agreement with the similarity of these equations, see Theorem~\ref{thm:GroupClassificationGroupoid}.
The compacton equations $K(n,n)$, where $f=g=nu^{n-1}$,
belong to Case~2 of Table~\ref{tab:CLChars}, and thus they possess, in addition to the conserved densities $u$ and $u^{n+1}$ (up to constant multipliers, with $u^{n+1}$ replaced by $\ln|u|$ for $n=-1$),
the conserved densities $u\cos x$ and $u\sin x$, as was noted in~\cite{rose1997a}.
The defocusing version $u_t-(u^n)_x+(u^n)_{xxx}=0$ of these equations belongs to Case~3 of Table~\ref{tab:CLChars}, and the corresponding additional conserved densities are $u\cosh x$ and $u\sinh x$.
Two similar conservation laws were also found in~\cite{rose1997a} for the equations $K(-1-n,-1-n)$ written in Lagrangian coordinates, but they are nonlocal for the equations $K(n+2,n)$ in the original variables.
For the KdV equation, Case~4 of Table~\ref{tab:CLChars} gives the Galilean conservation law with the characteristic $x-tu$.
Comparing Theorem~\ref{thm:CLs} with Theorem~\ref{thm:GroupClassification}, we see that
the equations with $f=0$ from Cases~1, 6, 7 and~9a of Table~\ref{tab:GroupClassification} belong to Case~1 or~5 of Table~\ref{tab:CLChars},
and the equations from Cases~9b and~9c of Table~\ref{tab:GroupClassification} belong to Cases~2 and~3 of Table~\ref{tab:CLChars}.
At the same time, there are equations with additional conservation laws with characteristics of nonpositive order,
e.g., those from Cases~2--4 of Table~\ref{tab:CLChars} with general~$g$, that admit no Lie-symmetry extensions.
The equations $u_t+a(u^n)_x+b(u^n)_{xxx}=0$ with $bn\ne0$ and $n\ne1$ considered in~\cite{vodo2013a}
are the equations from the class~\eqref{eq:GenKdV} with $g=bnu^{n-1}$ and $f=(a/b)g$,
and thus they are $G^\sim$-equivalent to equations from Cases~1--3 of Table~\ref{tab:CLChars} with power~$g$.
It was proved in~\cite{vodo2013a} that for $n\ne-2,-\frac12$ any conservation law of such an equation
has a conserved density depending at most on $(t,x,u)$ up to adding trivial conserved densities,
and the corresponding four conservation laws are exactly those given by Theorem~\ref{thm:CLs}.
The exceptional values $n=-2$ and $n=-\frac12$ correspond to the symmetry integrable equations,
and the value $n=-\frac12$ singles out the equations from the subclass~$\mathcal L_{\rm HD}$.
Therefore, Theorem~\ref{thm:CLs} extends the description of conservation laws with characteristics of nonpositive order
from~\cite{vodo2013a} to the entire class~\eqref{eq:GenKdV}.
\end{remark}

\section{Conclusion}\label{sec:Conclusion}

In the present paper, we have carried out the complete group classification of the class~\eqref{eq:GenKdV} of generalized KdV equations.
We have computed the equivalence algebra~$\mathfrak g^\sim$ of this class~\eqref{eq:GenKdV} by the infinitesimal method
and constructed its equivalence group~$G^\sim$ in two ways,
by the advanced version of the direct method and by the megaideal-based version of the algebraic method.
These two computations are independent of each other and thus provide a double check of the result.
Within the advanced version of the direct method, we have used known restrictions on admissible transformations of wider classes of evolution equations,
which essentially simplifies the computation.
Within the algebraic method, the main restrictions on equivalence transformations are derived from the invariance
of the megaideals $\mathfrak m_4$, \dots, $\mathfrak m_8$ of~$\mathfrak g^\sim$,
which are obtained by standard operations such as taking derived algebras, centers and centralizers,
and the direct computation is needed only in the last step for a much simpler form of transformations.
The group~$G^\sim$ is isomorphic to the subgroup of index two in ${\rm Aut}(\mathfrak g^\sim)$
that consists of the automorphisms with matrices of the form~\eqref{eq:AutContinuous}.
The automorphism~$\mathscr J$ of~$\mathfrak g^\sim$, which is an involution permuting $P^t=\p_t$ and $H=t\p_x+\p_f$, is not induced by an equivalence transformation.
The direct computation in the last step of the algebraic proof
is needed for relating the scaling of~$g$ to the scalings of the independent variables.
The group~$G^\sim$ is generated by the shifts, the scalings and the reflections of the variables and by the Galilean boosts,
and its quotient group with respect to its identity component is isomorphic to $\mathbb Z_2\times\mathbb Z_2\times\mathbb Z_2$.
Note that the set of megaideals of the seven-dimensional algebra~$\mathfrak g^\sim$ is infinite
since it contains the one-parameter family $\{\mathfrak m_{10}^c\}$.
The megaideals $\mathfrak m_9$ and~$\mathfrak m_{10}^c$ cannot be obtained by standard operations,
but they can be found without knowing the entire group ${\rm Aut}(\mathfrak g^\sim)$
by evaluating the defining property of automorphisms on certain pairs of elements of~$\mathfrak g^\sim$, see Lemma~\ref{lem:MegaidealOfg} and cf.~\cite{boyk2024a}.

Before carrying out the group classification of the class~\eqref{eq:GenKdV}, we have simplified the system of determining equations for Lie symmetries
and reduced it to the representation~\eqref{eq:DeterEqComponents} for the components of Lie-symmetry vector fields
and three classifying equations~\eqref{eq:SimplDeterEqI}--\eqref{eq:SimplDeterEqIII}.
For each Lie-symmetry vector field and each fixed point~$(t,x)$,
the first two classifying equations give a template-form system for the pair~$(f,g)$,
and the main cases of the furcate splitting are
singled out by the number~$k$ of such systems with linearly independent parameter tuples.
In view of the relations between the parameters of template-form systems that follow from their expressions in terms of the components of Lie symmetries,
we have $k\leqslant2$ for any equation from the class~\eqref{eq:GenKdV}.
The main cases $k=0$, $k=1$ and $k=2$ are split into subcases depending on the values of the parameters,
and these subcases are split further with the help of the third classifying equation~\eqref{eq:SimplDeterEqIII},
which also gives the dependence of the parameter function~$\Lambda$ on~$(t,x)$.
Combining the template-form systems with the relations~\eqref{eq:k1Rels} and~\eqref{eq:k1RelConseqs} between their parameters,
we have naturally singled out the case $\mu=3\nu$, $\alpha=-2\nu$, where $\Lambda$ may depend on~$x$
and the parameter~$\delta$ is essential.
This case leads to the subclass~$\mathcal L_{\rm HD}$, which is the only one with $\xi_{xx}\ne0$.
If $\eta^0=0$ or $\eta^1=0$, i.e., $\eta=\zeta u$ or $\eta=\zeta$ for a function $\zeta=\zeta(t,x)$,
then the third classifying equation reduces to $\zeta_xf+\zeta_{xxx}g+\zeta_t=0$.
For each fixed~$(t,x)$, this is a template-form system of degenerate type, with $\alpha=\beta=\mu=0$,
where the first equation is an identity and the second one is the algebraic equation $\nu f+\delta g+\kappa=0$.
In the case $\mu=3\nu$, $\alpha=-2\nu$ with $\nu\ne0$, where $\eta=\Lambda u$ after the gauge,
the splitting of this equation separates the subclass~$\mathcal L_{\rm HD}$ from Case~2.
For $k=2$, if the projection $\vartheta\mapsto(\alpha,\beta)$ on~$V_\theta$ is not surjective, a degenerate template-form system arises from its kernel of the projection $\vartheta\mapsto(\alpha,\beta)$ on~$V_\theta$,
and it implies that $f$ is proportional to~$g$ up to the $G^\sim$-equivalence.
For classes with several arbitrary elements depending on the same argument,
the classifying equations of template form should be considered jointly as a system,
and degenerate template-form systems, which may also be given by the other classifying equations,
are essential at the further steps of the furcate splitting.

The maximal Lie invariance algebras of equations from the class~\eqref{eq:GenKdV} are of dimensions two, three, four or five.
The dimension five is attained only for the equations from the subclass~$\mathcal L_{\rm HD}$,
whose maximal Lie invariance algebras are isomorphic to $\mathrm{sl}(2,\mathbb R)\oplus A_{2.1}$,
and the dimension four is attained only for the equations that are $G^\sim$-equivalent to
the equations with $f=0$ and $g=|u|^\mu$, $\mu\ne0,-\frac32$, or $g=e^u$, and to the KdV equation.
All the equations from~$\mathcal L_{\rm HD}$ are similar to the Harry Dym equation,
and the point transformations between Cases~9a, 9b and~9c are the only additional point equivalence transformations
between the listed $G^\sim$-inequivalent classification cases.
The algebras of Cases~9a--9c are isomorphic, but their realizations are not related by transformations from~$G^\sim$,
which is consistent with the fact that the corresponding point transformations are not induced by elements of~$G^\sim$.
In the terminology of~\cite{vane2020b}, the Lie-symmetry extensions for the equations from the subclass~$\mathcal L_{\rm gen}$, Cases~1--8,
are regular, i.e., they are the projections of subalgebras of the equivalence algebra~$\mathfrak g^\sim$,
whereas the Lie-symmetry extensions for the subclass~$\mathcal L_{\rm HD}$, Cases~9a--9c, are singular.
Therefore, the Lie-symmetry extensions of the maximal dimension five
cannot be obtained by the classical algebraic method of group classification,
which is why we use the method of furcate splitting.

We have also classified the conservation laws of equations from the class~\eqref{eq:GenKdV} with characteristics of nonpositive order
and presented conserved currents contained in these conservation laws.
Up to the $G^\sim$-equivalence, there are seven cases of extensions of the space of such characteristics,
which generalizes the results of~\cite{vodo2013a} on the equations of $K(n,n)$ type to the entire class~\eqref{eq:GenKdV}.
In particular, this space is four-dimensional for all equations from the subclass~$\mathcal L_{\rm HD}$.
For the focusing and defocusing versions of the compacton equations $K(n,n)$, the additional conserved densities are $u\cos x$, $u\sin x$, cf.~\cite{rose1997a}, and $u\cosh x$, $u\sinh x$, respectively.
At the same time, this classification does not follow the group classification:
there are equations with additional conservation laws of this kind that admit no Lie-symmetry extensions,
e.g., those from Cases~2--4 of Table~\ref{tab:CLChars} with general~$g$, and thus the existence of such conservation laws is not related to an extension of the Lie invariance algebra.

The results of the present paper provide a basis for the further study of equations from the class~\eqref{eq:GenKdV}
within the framework of group analysis of differential equations,
including Lie reductions and exact solutions of equations with Lie-symmetry extensions, cf.~\cite{vinn2024a,vinn2026a},
the classification of conservation laws of higher orders and potential symmetries~\cite{kunz2008a},
and the group classification of wider classes of equations with nonlinear dispersion.
Another problem for further study is the description of the entire equivalence groupoid of the class~\eqref{eq:GenKdV}.
Since the transformations~\eqref{eq:HDMappingB} and~\eqref{eq:HDMappingC} relate $G^\sim$-inequivalent equations,
the class~\eqref{eq:GenKdV} is not normalized and even not semi-normalized.
The subclasses~$\mathcal L_{\rm gen}$ and~$\mathcal L_{\rm HD}$ are not related by point transformations
since such transformations preserve the dimension of maximal Lie invariance algebras,
$\dim\mathfrak g_\theta=5$ for each equation from~$\mathcal L_{\rm HD}$ and $\dim\mathfrak g_\theta\leqslant4$ for each equation from~$\mathcal L_{\rm gen}$.
A preliminary consideration has shown that the subclass~$\mathcal L_{\rm gen}$ is normalized in the usual sense~\cite{popo2010a,vane2020b} with respect to the equivalence group~$G^\sim$,
i.e., the only obstacle to the normalization of the class~\eqref{eq:GenKdV} is the subclass~$\mathcal L_{\rm HD}$.
This would underlie the fact that the Lie-symmetry extensions for equations from~$\mathcal L_{\rm gen}$, Cases~1--8 of Table~\ref{tab:GroupClassification}, are regular,
whereas the singular Cases~9a--9c are related to the subclass~$\mathcal L_{\rm HD}$.

\appendix

\section{Automorphisms and megaideals of the equivalence algebra}\label{sec:AutAndMegaideals}

In this appendix, we construct the automorphism group ${\rm Aut}(\mathfrak g^\sim)$ of the equivalence algebra~$\mathfrak g^\sim$
of the class~\eqref{eq:GenKdV} and describe all megaideals of~$\mathfrak g^\sim$.
We use the basis $(e_1,\dots,e_7)=(P^x,P^t,H,D^t,D^x,P^u,D^u)$ of~$\mathfrak g^\sim$
and identify each linear operator~$\mathscr A$ on~$\mathfrak g^\sim$ with its matrix $(a_{ij})$ in this basis,
$\mathscr Ae_j=\sum_{i=1}^7a_{ij}e_i$, $j=1,\dots,7$,
i.e., the $j$th column of the matrix consists of the components of~$\mathscr Ae_j$.

The automorphism group ${\rm Aut}(\mathfrak g^\sim)$ can be computed in two ways.
Within the direct method, the condition that a nondegenerate linear operator~$\mathscr A$ on~$\mathfrak g^\sim$ is an automorphism of~$\mathfrak g^\sim$,
\[
[\mathscr Ae_i,\mathscr Ae_j]=\mathscr A[e_i,e_j],\quad i,j=1,\dots,7,
\]
is a system of quadratic equations for the 49 unknown entries~$a_{ij}$ of its matrix,
which can be solved, e.g., using {\sf Maple}.
Within the megaideal-based method, the megaideals $\mathfrak m_4$, \dots, $\mathfrak m_8$ constructed in the algebraic proof of Theorem~\ref{thm:EquivGroup}
reduce the number of unknown entries to 22 and immediately single out the block structure of automorphisms.
Both methods lead to the following result.

\begin{proposition}\label{pro:AutGroup}
The automorphism group ${\rm Aut}(\mathfrak g^\sim)$ of the equivalence algebra~$\mathfrak g^\sim$
consists of the linear operators on~$\mathfrak g^\sim$ whose matrices are of the form
\begin{gather}\label{eq:AutContinuous}
\begin{pmatrix}
a_{22}a_{33} & -a_{22}a_{35} & a_{24}a_{33} & -a_{24}a_{35} & a_{15} & 0 & 0\\
0 & a_{22} & 0 & a_{24} & 0 & 0 & 0\\
0 & 0 & a_{33} & -a_{35} & a_{35} & 0 & 0\\
0 & 0 & 0 & 1 & 0 & 0 & 0\\
0 & 0 & 0 & 0 & 1 & 0 & 0\\
0 & 0 & 0 & 0 & 0 & a_{66} & a_{67}\\
0 & 0 & 0 & 0 & 0 & 0 & 1
\end{pmatrix},
\quad a_{22}a_{33}a_{66}\ne0,
\\[1ex] \label{eq:AutSecondComponent}
\begin{pmatrix}
-a_{23}a_{32} & a_{25}a_{32} & -a_{23}a_{34} & a_{25}a_{34} & a_{15} & 0 & 0\\
0 & 0 & a_{23} & -a_{25} & a_{25} & 0 & 0\\
0 & a_{32} & 0 & a_{34} & 0 & 0 & 0\\
0 & 0 & 0 & -1 & 1 & 0 & 0\\
0 & 0 & 0 & 0 & 1 & 0 & 0\\
0 & 0 & 0 & 0 & 0 & a_{66} & a_{67}\\
0 & 0 & 0 & 0 & 0 & 0 & 1
\end{pmatrix},
\quad a_{23}a_{32}a_{66}\ne0,
\end{gather}
where the parameters~$a_{ij}$ are arbitrary constants satisfying the indicated inequalities.
The automorphisms with matrices of the form~\eqref{eq:AutContinuous} constitute a subgroup of index two in~${\rm Aut}(\mathfrak g^\sim)$.
The matrix of the form~\eqref{eq:AutSecondComponent} with $a_{23}=a_{32}=a_{66}=1$ and $a_{15}=a_{25}=a_{34}=a_{67}=0$
corresponds to the involution
\begin{gather}\label{eq:AutDiscrete}
\begin{split}
\mathscr J\colon\quad&
P^x\mapsto-P^x,\quad P^t\mapsto H,\quad H\mapsto P^t,\quad D^t\mapsto-D^t,\\&
D^x\mapsto D^t+D^x,\quad P^u\mapsto P^u,\quad D^u\mapsto D^u,
\end{split}
\end{gather}
and each automorphism with matrix of the form~\eqref{eq:AutSecondComponent}
is the composition $\mathscr A\circ\mathscr J$ of~$\mathscr J$ with an automorphism~$\mathscr A$ with matrix of the form~\eqref{eq:AutContinuous}.
In other words, each matrix of the form~\eqref{eq:AutSecondComponent} is the product of a matrix of the form~\eqref{eq:AutContinuous}
and the matrix of~$\mathscr J$, which is the block-diagonal matrix
\[
(-1)\oplus\begin{pmatrix}0&1\\1&0\end{pmatrix}\oplus\begin{pmatrix}-1&1\\0&1\end{pmatrix}\oplus\begin{pmatrix}1&0\\0&1\end{pmatrix}.
\]
\end{proposition}

\begin{proof}
Let $\mathscr A\in{\rm Aut}(\mathfrak g^\sim)$.
Since the subspaces $\mathfrak m_4$, \dots, $\mathfrak m_8$ are megaideals of~$\mathfrak g^\sim$, we have
\[
\mathscr AP^x\in\mathfrak m_7,\quad \mathscr AP^u\in\mathfrak m_8,\quad \mathscr AD^u\in\mathfrak m_5,\quad
\mathscr AP^t,\ \mathscr AH\in\mathfrak m_4,\quad \mathscr AD^t\in\mathfrak m_6.
\]
Therefore, only 22 of 49 entries of the matrix of~$\mathscr A$ may be nonzero.
Hence $\mathscr AD^t=rD^t+Y$ with a nonzero constant~$r$ and $Y\in\mathfrak m_4$.
As $[\mathfrak m_4,\mathfrak m_4]\subseteq\mathfrak m_7$,
the operator ${\rm ad}_{\mathscr AD^t}$ induces the operator $r\,{\rm ad}_{D^t}$ on the quotient space $\mathfrak m_4/\mathfrak m_7$,
whose eigenvectors are the cosets of $P^t$ and~$H$ with the eigenvalues $-r$ and~$r$, respectively.
The equalities $[\mathscr AP^t,\mathscr AD^t]=\mathscr AP^t$ and $[\mathscr AH,\mathscr AD^t]=-\mathscr AH$ then imply that either $r=1$,
and $\mathscr AP^t\in\langle P^t,P^x\rangle$ and $\mathscr AH\in\langle H,P^x\rangle$,
or $r=-1$, and $\mathscr AP^t\in\langle H,P^x\rangle$ and $\mathscr AH\in\langle P^t,P^x\rangle$.
In the first case, the remaining commutation relations~\eqref{eq:CommRelations} are satisfied
if and only if the matrix of~$\mathscr A$ is of the form~\eqref{eq:AutContinuous}.
It is straightforward to check that the mapping~\eqref{eq:AutDiscrete} is an automorphism of~$\mathfrak g^\sim$.
In the second case, the automorphism $\mathscr A\circ\mathscr J$ belongs to the first case,
and thus $\mathscr A=(\mathscr A\circ\mathscr J)\circ\mathscr J$ has the matrix of the form~\eqref{eq:AutSecondComponent}.
\end{proof}

In terms of the vector fields spanning~$\mathfrak g^\sim$, the automorphisms with matrices of the form~\eqref{eq:AutContinuous} are
\begin{gather*}
P^x\mapsto a_{22}a_{33}P^x,\quad
P^t\mapsto a_{22}(P^t-a_{35}P^x),\quad
H\mapsto a_{33}(H+a_{24}P^x),\quad
P^u\mapsto a_{66}P^u,\\
D^t\mapsto D^t+a_{24}(P^t-a_{35}P^x)-a_{35}H,\quad
D^x\mapsto D^x+a_{35}H+a_{15}P^x,\quad
D^u\mapsto D^u+a_{67}P^u.
\end{gather*}
The one-parameter subgroups of these automorphisms that are parameterized by~$a_{35}$, $a_{24}$, $a_{15}$ and~$a_{67}$
consist of the inner automorphisms generated by~$H$, $P^t$, $P^x$ and~$P^u$, respectively.

\begin{remark}\label{rem:AutGroup}
The pushforwards of vector fields by the transformations~\eqref{eq:EquivGroupTransform}
are exactly the automorphisms with matrices of the form~\eqref{eq:AutContinuous}, where
\[
a_{22}=c_1,\quad a_{33}=\frac{c_2}{c_1},\quad a_{66}=c_5,\quad a_{35}=-\frac{c_3}{c_1},\quad a_{24}=-c_0,\quad
a_{15}=\frac{c_0c_3}{c_1}-c_4,\quad a_{67}=-c_6,
\]
and this correspondence is one-to-one.
Therefore, the group~$G^\sim$ is isomorphic to a subgroup of index two in ${\rm Aut}(\mathfrak g^\sim)$.
The involution~$\mathscr J$, which interchanges $P^t=\p_t$ and $H=t\p_x+\p_f$, is not induced by any point equivalence transformation of the class~\eqref{eq:GenKdV}.
Indeed, in the algebraic proof of Theorem~\ref{thm:EquivGroup}, the conditions for~$\mathfrak m_7$ and~$\mathfrak m_8$ give $T=T(t)$,
and thus the pushforward $\Psi_*P^t$ has the nonzero component $T_t$ along~$\p_{\tilde t}$, whereas $H$ has no such component.
This is why not every automorphism of~$\mathfrak g^\sim$ is induced by an equivalence transformation.
The direct computation in the last step of the algebraic proof is needed for deriving the relation $\lambda_0=a_{11}^{\,3}a_{22}^{-1}$.
\end{remark}

Knowing all automorphisms of a Lie algebra, one can find all its megaideals
since megaideals are exactly the subspaces that are invariant with respect to all automorphisms.
In this way, we have found all megaideals of the algebra~$\mathfrak g^\sim$ using {\sf Maple}.
In addition to the megaideals $\mathfrak m_1$, \dots, $\mathfrak m_8$
constructed in the algebraic proof of Theorem~\ref{thm:EquivGroup}, they include the megaideals
\begin{gather*}
\mathfrak m_9:=\langle P^x,P^t,H,D^t+2D^x\rangle,\quad
\mathfrak m_{10}^c:=\langle P^x,P^t,H,P^u,D^t+2D^x+cD^u\rangle,\\
\mathfrak m_{11}:={\rm C}(\mathfrak m_9)=\langle P^u,D^u\rangle,
\end{gather*}
where $c$ is an arbitrary constant.
Following~\cite{boyk2024a}, we call a proper megaideal of~$\mathfrak g^\sim$ \emph{essential} if it is not the sum of other proper megaideals of~$\mathfrak g^\sim$.

The megaideals $\mathfrak m_9$ and~$\mathfrak m_{10}^c$ are not obtained from the megaideals $\mathfrak m_1$, \dots, $\mathfrak m_8$ by standard operations
such as taking derived algebras, centers, centralizers, sums and intersections.
Nevertheless, they can also be found without knowing the entire group ${\rm Aut}(\mathfrak g^\sim)$,
by evaluating the defining property of automorphisms on certain pairs of elements of~$\mathfrak g^\sim$,
which is similar to the technique for finding the megaideal~$\bar{\mathfrak m}_7$ in~\cite[Lemma~10]{boyk2024a}.

\begin{lemma}\label{lem:MegaidealOfg}
The subspaces $\mathfrak m_9$, $\mathfrak m_{10}^c$, where $c$ is an arbitrary constant, and~$\mathfrak m_{11}$ are megaideals of~$\mathfrak g^\sim$.
\end{lemma}

\begin{proof}
Let $\mathscr A\in{\rm Aut}(\mathfrak g^\sim)$.
We use the matrix notation $\mathscr Ae_j=\sum_ia_{ij}e_i$ introduced at the beginning of this appendix.
Since $\mathfrak m_4$, $\mathfrak m_5$, $\mathfrak m_6$, $\mathfrak m_7$, $\mathfrak m_8$ and ${\rm C}(\mathfrak m_8)=\langle P^x,P^t,H,D^t,D^x,P^u\rangle$ are megaideals of~$\mathfrak g^\sim$,
we have
\begin{gather*}
\mathscr AP^x=a_{11}P^x,\quad \mathscr AP^u=a_{66}P^u,\quad \mathscr AD^u=a_{77}D^u+a_{17}P^x+a_{67}P^u,\quad \mathscr AP^t,\ \mathscr AH\in\mathfrak m_4,\\
\mathscr AD^t=a_{44}D^t+Y_1,\quad \mathscr AD^x=a_{45}D^t+a_{55}D^x+Y_2+a_{65}P^u,
\end{gather*}
where $a_{ij}$ are constants with $a_{11}a_{44}a_{66}\ne0$, and $Y_1,Y_2\in\mathfrak m_4$.
The equalities $[\mathscr AP^u,\mathscr AD^u]=\mathscr AP^u$ and $[\mathscr AP^x,\mathscr AD^x]=\mathscr AP^x$ give $a_{77}=1$ and $a_{55}=1$.
The operator ${\rm ad}_{\mathscr AD^t}$ induces the operator $a_{44}{\rm ad}_{D^t}$ on the quotient space $\mathfrak m_4/\mathfrak m_7$,
whose eigenvectors are the cosets of $P^t$ and~$H$ with the eigenvalues $-a_{44}$ and~$a_{44}$, respectively.
Hence the equalities $[\mathscr AP^t,\mathscr AD^t]=\mathscr AP^t$ and $[\mathscr AH,\mathscr AD^t]=-\mathscr AH$ imply that
either $a_{44}=1$ and $\mathscr AP^t\in\langle P^t,P^x\rangle$, or $a_{44}=-1$ and $\mathscr AP^t\in\langle H,P^x\rangle$.
Evaluating the equality $[\mathscr AP^t,\mathscr AD^x]=\mathscr A[P^t,D^x]=0$ modulo~$\mathfrak m_7$,
we derive $a_{45}=0$ if $a_{44}=1$ and $a_{45}=1$ if $a_{44}=-1$, i.e., $a_{44}+2a_{45}=1$ in both cases.
Finally, the equality $[\mathscr AD^u,\mathscr AD^x]=\mathscr A[D^u,D^x]=0$ gives $a_{17}P^x-a_{65}P^u=0$, i.e., $a_{17}=a_{65}=0$.
Therefore,
\[
\mathscr A(D^t+2D^x)=D^t+2D^x+Y_1+2Y_2,\qquad \mathscr AD^u=D^u+a_{67}P^u,
\]
and thus $\mathscr A\mathfrak m_9\subseteq\mathfrak m_9$ and $\mathscr A\mathfrak m_{10}^c\subseteq\mathfrak m_{10}^c$ for any automorphism~$\mathscr A$ of~$\mathfrak g^\sim$.
Moreover, $\mathfrak m_{11}={\rm C}(\mathfrak m_9)$.
\end{proof}

\begin{proposition}\label{pro:AllMegaideals}
The megaideals $\mathfrak m_4$, $\mathfrak m_6$, $\mathfrak m_7$, $\mathfrak m_8$, $\mathfrak m_9$, $\mathfrak m_{10}^c$ with $c\ne0$ and~$\mathfrak m_{11}$
exhaust the essential megaideals of~$\mathfrak g^\sim$,
and each megaideal of~$\mathfrak g^\sim$ is the sum of some of them.
More specifically, the complete list of nonzero megaideals of~$\mathfrak g^\sim$ consists of the above essential megaideals and
\begin{gather*}
\mathfrak m_4+\mathfrak m_8=\mathfrak m_1,\quad
\mathfrak m_7+\mathfrak m_8=\mathfrak m_2,\quad
\mathfrak m_6+\mathfrak m_8=\mathfrak m_3,\quad
\mathfrak m_7+\mathfrak m_{11}=\mathfrak m_5,\quad
\mathfrak m_9+\mathfrak m_8=\mathfrak m_{10}^0,\\
\mathfrak m_4+\mathfrak m_{11},\quad
\mathfrak m_6+\mathfrak m_{11}={\rm C}(\mathfrak m_7),\quad
\mathfrak m_9+\mathfrak m_{11},\quad
\mathfrak m_6+\mathfrak m_9,\\
\mathfrak m_6+\mathfrak m_{10}^c=\langle P^x,P^t,H,P^u,D^t,D^x+\tfrac c2D^u\rangle,\quad
\mathfrak g^\sim=\mathfrak m_6+\mathfrak m_9+\mathfrak m_{11},
\end{gather*}
where $c\in\mathbb R$, and, in particular, $\mathfrak m_6+\mathfrak m_{10}^0={\rm C}(\mathfrak m_8)$.
\end{proposition}

\begin{proof}
A subspace of~$\mathfrak g^\sim$ is a megaideal if and only if it is invariant with respect to the automorphisms
with matrices of the form~\eqref{eq:AutContinuous} and the involution~$\mathscr J$.
Invariance with respect to the automorphisms~\eqref{eq:AutContinuous} with $a_{15}=a_{24}=a_{35}=a_{67}=0$
means that the subspace is the sum of its intersections with the weight subspaces
$\langle P^x\rangle$, $\langle P^t\rangle$, $\langle H\rangle$, $\langle P^u\rangle$ and $\langle D^t,D^x,D^u\rangle$.
The invariance with respect to the one-parameter subgroups of automorphisms~\eqref{eq:AutContinuous}
parameterized by~$a_{35}$, $a_{24}$, $a_{15}$ and~$a_{67}$ is equivalent to the invariance with respect to the inner derivations
${\rm ad}_H$, ${\rm ad}_{P^t}$, ${\rm ad}_{P^x}$ and~${\rm ad}_{P^u}$, respectively.
Hence, if a megaideal~$\mathfrak m$ contains $aD^t+bD^x+cD^u$, then it contains $(b-a)H$, $aP^t$, $bP^x$ and $cP^u$.
Moreover, $P^t\in\mathfrak m$ if and only if $H\in\mathfrak m$ in view of the invariance with respect to~$\mathscr J$, and then $P^x\in\mathfrak m$.
The restriction of~$\mathscr J$ to $\langle D^t,D^x,D^u\rangle$ has the eigenspaces
$\langle D^t+2D^x,D^u\rangle$ and $\langle D^t\rangle$ with the eigenvalues~$1$ and~$-1$, respectively,
and thus the intersection of~$\mathfrak m$ with $\langle D^t,D^x,D^u\rangle$ is the sum of its intersections with these eigenspaces.
Running through all possibilities, we obtain the above list.
The megaideals that are not listed as sums are essential
since, for each of them, the sum of the megaideals that are properly contained in it is a proper subspace of it.
\end{proof}

\begin{remark}\label{rem:Megaideals}
The family $\{\mathfrak m_{10}^c\}$ shows that the set of megaideals of a finite-dimensional Lie algebra may be infinite.
Note that megaideals of real Lie algebras of dimension up to four do not involve continuous parameters,
and thus the number of megaideals of each of these algebras is finite~\cite{popo2003a}.
The proper megaideals $\mathfrak m_1$, $\mathfrak m_2$, $\mathfrak m_3$ and~$\mathfrak m_5$ are inessential.
The megaideals $\mathfrak m_1$, $\mathfrak m_2$ and~$\mathfrak m_3$ are used only for constructing the other megaideals,
whereas the megaideal~$\mathfrak m_5$ is used in the algebraic proof of Theorem~\ref{thm:EquivGroup}
and in the proof of Proposition~\ref{pro:AutGroup} since, unlike the essential megaideal~$\mathfrak m_{11}\subset\mathfrak m_5$,
it is obtained by standard operations.
\end{remark}

\section*{Acknowledgments}
\begin{table}[!ht]
\begin{minipage}{20mm}
    \includegraphics[width=20mm]{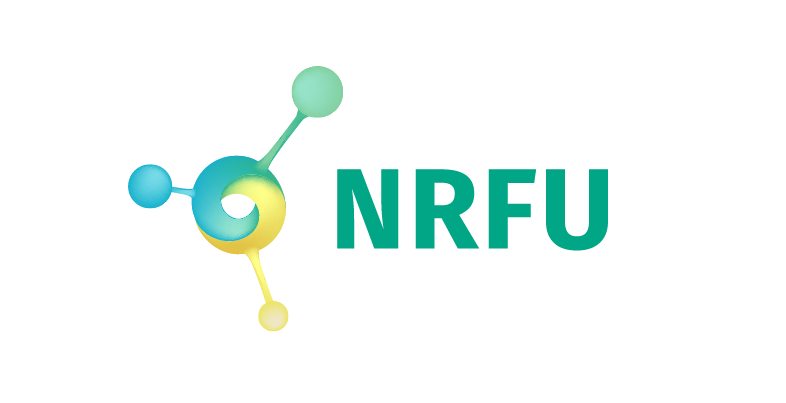}
\end{minipage}
\
\begin{minipage}{138mm}
O.O.V. expresses gratitude to the National Research Foundation of Ukraine, thanks to whose grant support the project 2025.07/0405 ``Algebraic methods for studying equations of mathematical physics'' is being implemented.
The contents of this work do not necessarily reflect the views of the National Research Foundation of Ukraine and are the sole responsibility of the Institute of Mathematics of NAS of Ukraine.\looseness=-1
\end{minipage}
\end{table}
\vspace{-4.2mm}
\noindent
The author is grateful to her supervisors Roman Popovych and Vyacheslav Boyko for posing the problem, valuable discussions and constant support.
During the preparation of this manuscript, the author used Claude (Anthropic) for language editing and for checking the consistency of notation and cross-references.
All the results, computations and proofs are the author's own.
The author reviewed all the suggested changes and takes full responsibility for the content of the paper.
The author expresses her deepest thanks to the Armed Forces of Ukraine and the civilian people of Ukraine
for their bravery and courage in defense of peace and freedom in Europe and in the entire world from russism.


\end{document}